\documentclass[11pt,reqno]{amsart}
\usepackage[latin9]{inputenc}
\usepackage{amstext}
\usepackage{amsthm}
\usepackage{amssymb}
\usepackage{graphicx}
\usepackage{comment}
\usepackage{amsmath}
\usepackage{breqn}
\usepackage[pagewise]{lineno}
\makeatletter
\numberwithin{equation}{section}
\numberwithin{figure}{section}
\theoremstyle{plain}

 \theoremstyle{remark}

\newtheorem{theorem}{Theorem}[section]

\newtheorem{lemma}[theorem]{Lemma}

\newtheorem{remark}[theorem]{Remark}

\usepackage{amsthm}
\usepackage{color}
\usepackage{color}\usepackage{xcolor}
\usepackage{subcaption}
\usepackage{enumerate}
\usepackage[draft]{todonotes}
\usepackage[american]{babel}
\usepackage{geometry}

\usepackage{color,soul}

\usepackage[colorlinks,linkcolor=red,citecolor=blue,urlcolor=blue]{hyperref}

\newcommand{\ri}{\mathrm{i}}

\newcommand{\tr}{\mathrm{trace}}
\newcommand{\erf}{\mathrm{erf}}
\newcommand{\sign}{\mathrm{sign}}
\newcommand{\erfi}{\mathrm{erfi}}
\newcommand{\range}{\mathrm{range}}
\newcommand{\codim}{\mathrm{codim}}
\newcommand{\ind}{\mathrm{ind}}

\def\Id {{ \rm Id}}

\makeatother

\providecommand{\remarkname}{Remark}
\providecommand{\theoremname}{Theorem}

\begin{document}

\title[Spectrum BGK]{Exact Hydrodynamic Manifolds for the Linear Boltzmann BGK Equation I: Spectral Theory}

\author{Florian Kogelbauer}
\address{ETH Z\"{u}rich, Department of Mechanical and Process Engineering, Leonhardstrasse 27, 8092 Z\"{u}rich, Switzerland}
\email{floriank@ethz.ch}

\author{Ilya Karlin}
\address{ETH Z\"{u}rich, Department of Mechanical and Process Engineering, Leonhardstrasse 27, 8092 Z\"{u}rich, Switzerland}
\email{ikarlin@ethz.ch}

\begin{abstract}
We perform a complete spectral analysis of the linear three-dimensional Boltzmann BGK operator resulting in an explicit transcendental equation for the eigenvalues. Using the theory of finite-rank perturbations, we confirm the existence of a critical wave number $k_{\rm crit}$ which limits the number of hydrodynamic modes in the frequency space. This implies that there are only finitely many isolated eigenvalues above the essential spectrum at each wave number, thus showing the existence of a finite-dimensional, well-separated linear hydrodynamic manifold as a combination of invariant eigenspaces. The obtained results can serve as a benchmark for validating approximate theories of hydrodynamic closures and moment methods and provides the basis for the spectral closure operator. 
\end{abstract}

\maketitle

\section{Introduction}

The derivation of hydrodynamic equations from kinetic theory is a fundamental, yet not completely resolved, problem in thermodynamics and fluids, dating back at least to part (b) of Hilbert's sixth problem \cite{hilbert2000mathematical}. Given the Boltzmann equation or an approximation of it, can the the basic equations of fluid dynamics (Euler, Navier--Stokes) be derived directly from the dynamics of the distribution function?\\

One classical approach is to seek a series expansion in terms of a small parameter, such as the relaxation time $\tau$ or the Knudsen number $\varepsilon$ \cite{truesdell1980fundamentals}. One widely used expansion is the Chapman--Enskog series \cite{chapman1990mathematical}, where it is assumed that the collision term scales with $\varepsilon^{-1}$, thus indicating a (singular) Taylor expansion in $\varepsilon$. Indeed, the zeroth order PDE obtained this way gives the Euler equation, while the first order PDE reproduces the Navier--Stokes equation. On the linear level, the Navier--Stokes equation is globally dissipative and decay of entropy on the kinetic level translates to decay of energy on the fluid level.\\

For higher-order expansions, however, we are in trouble. In \cite{bobylev1982chapman}, it was first shown that an expansion in terms of Knudsen number can lead to nonphysical properties of the hydrodynamic models: At order two (Burnett equation \cite{chapman1990mathematical}), the dispersion relation shows a change of sign, thus leading to modes which grow in energy (Bobylev instability). In particular, the Burnett hydrodynamics are not hyperbolic and there exists no H-theorem for them \cite{bobylev2006instabilities}.\\

From a mathematical point of view, of course, there is no guarantee that the expansion of a non-local operator in frequency space, i.e., an approximation in terms of local (differential) operators, gives a good approximation for the long-time dynamics of the overall system. Among the first to suggest a non-local closure relation was probably Rosenau \cite{PhysRevA.40.7193}. In a series of works (see, e.g., \cite{gorban1994method,gorbankarlin2013,gorban2005invariant} and references therein),
Karlin and Gorban derived explicit non-local closures by essentially summing the Chapman--Enskog series for all orders. Furthermore, we note that the Chapman--Enskog expansion mixes linear and nonlinear terms for the full Boltzmann equation since it only considers powers of $\varepsilon$, while the existence (and approximation) of a hydrodynamic manifold can be performed independently of the Knudsen number, for which it only enters as a parameter.\\

Spectral properties of linearized kinetic equations are of basic interest in thermodynamics and have been performed by numerous authors. Already Hilbert himself was concerned with the spectral properties of linear integral operators derived from the Boltzmann equation \cite{hilbert1912grundzuge}. Carleman \cite{carleman1957problemes} proved that the essential spectrum remains the same under a compact perturbation (Weyl's theorem) in the hard sphere case and was able to estimate the spectral gap. This result was generalized to a broader class of collision kernels by Grad \cite{grad1963asymptotic} and to soft potentials in \cite{caflisch1980boltzmann}.\\

For spatially uniform Maxwell molecules, a complete spectral description was derived in \cite{Bobylev1988} (together with exact special solutions and normal form calculations for the full, non-linear problem), see also \cite{chang1970studies}. Famously, in \cite{ellis1975first}, the fundamental properties of the spectrum of a comparably broad class of kinetic operators was derived in the small wave-number regime. In particular, the existence of eigenvalue branches and asymptotic expansion of the (small) eigenvalues for vanishing wave number was derived. This was carried further in, e.g., \cite{dudynski2013}.  \\

Let us also comment on the relation to Hilbert's sixth problem. Along these lines, several result on the convergence to the Navier--Stokes (and Euler) equations have been obtained. Already Grad \cite{grad1964asymptotic} was interested in this question. In \cite{ellis1975first}, it is also shown that the semi-group generated by the linearized Euler equation converges - for fixed time - to the semi-group generated by the linearized Boltzmann equation (and similarly, for the linear Navier--Stokes semi-group).
In \cite{saint2002discrete}, convergence of \textit{scaled solutions} to the Navier--Stokes equation along the lines of \cite{bardos1993fluid} was proved. We also mention the results related to convergence rates to the equilibrium (\textit{hypercoercivity}) of the variants of the BGK equation \cite{villani2006hypocoercivity,Desvillettes2010}. For an excellent review on the mathematical perspective of Hilbert's sixth problem, we refer to \cite{saint2014mathematical}.\\

In this work, we perform an explicit spectral analysis for the Bhatnagar--Gross--Krook (BGK) equation \cite{bhatnagar1954model} linearized around a global Maxwellian. The BGK model - despite being a comparatively simple approximation to the full Boltzmann equation - shares important features such as decay of entropy and the conservation laws of mass, momentum and energy \cite{bhatnagar1954model}. Global existence and estimates of the solution were proved in \cite{perthame1989global,perthame1993weighted} for the full, non-linear BGK system. For the spectral analysis of the BGK and related models, we refer to \cite{cercignani1988boltzmann}, where the curves of eigenvalues are described through a temporal Laplace transform. This technique was already used in, e.g., \cite{mason1970weak} to numerically evaluate the discrete eigenvalues of the BGK equation in the s-plane and to determine their limiting curves. In this work, we calculate these quadratures explicitly using a different technique - namely finite rank perturbations and the resolvent formalism. This will serve as a basis for the explicit calculation of the spectral closure performed in a subsequent work \cite{kogelbauerBGKspectral2}, based on the general theory detailed in \cite{kogelbauer2023rigorous}.\\

The single relaxation time $\tau$ in the BGK equation will serve as the analog of the Knudsen number and fundamental parameter in our analysis. Previous work on the full spectrum of kinetic models together with a hydrodynamic interpretation has been performed in \cite{KogelBoltz} for the three-dimensional Grad system, in \cite{kogelbauer2021} for the linear BGK equation with mass density only and in \cite{kogelbauer2024spectral} for the linear Shakhov model. A similar independent analysis for the one-dimensional linear BGK equation with one fluid moment was performed in \cite{ThomasCarty2017,carty2017elementary} in the context of grossly determined solutions (in the sense of \cite{truesdell1980fundamentals}) and recently in \cite{sukhtayev2024spectral} where convergence to the slow manifold is also proven in Sobolev space explicitly.\\

Indeed, we will give a complete and (up to the solution of a transcendental equation) explicit description of the spectrum of the BGK equation linearized around a global Maxwellian. We will confirm the existence of finitely many discrete eigenvalues above the essential spectrum \cite{cercignani1988boltzmann} as well as the existence of a critical wave number for each family of modes as zeros of an analytic function. More precisely, we show the following:
\begin{theorem}
The spectrum of the non-dimensional linearized BGK operator $\mathcal{L}$ (defined either on the real three space or on a three-dimensional torus) with relaxation time $\tau$ around a global Maxwellian is given by
\begin{equation}
\sigma(\mathcal{L}) = \left\{-\frac{1}{\tau}+\ri\mathbb{R}\right\}\cup\bigcup_{N\in \text{Modes}}\bigcup_{ k<k_{\rm crit,N}}\{\lambda_{N}(\tau k)\},
\end{equation}
where $\text{Modes}=\{\text{shear}, \text{diff}, \text{ac}, \text{ac}*\}$ corresponding to the shear mode, the diffusion mode and the pair of complex conjugate acoustic modes. The essential spectrum is given by the line $\Re\lambda=-\frac{1}{\tau}$, while the discrete spectrum at each wave number consists of a finite number of discrete, isolated eigenvalues. Along with each family of modes, there exists a critical wave number $k_{\rm crit,N}$, limiting the range of wave numbers (either continuous or discrete) for which $\lambda_N$ exists. 
\end{theorem}
While the proof in \cite{cercignani1988boltzmann} relies on a Laplace transform in space, our proof is based on the theory of finite-rank perturbations and the Weinstein--Aronszajn determinant, see e.g. \cite{kato1995perturbation} and \cite{aronszajn1948rayleigh,WEINSTEIN1974604}, together with some properties of the plasma dispersion function. We also refer to \cite{ljance1970completely} for a general discussion of regular perturbations of continuous spectra.\\

The paper is structured as follows: In Section \ref{definitions}, we introduce some notation and give some basic definitions. In Section \ref{Prelim}, we formulate the fundamental equations. Section \ref{spectralan} is devoted to the spectral analysis of the linear part, including the derivation of a spectral function describing the discrete spectrum completely. We also give a proof of the finiteness of the hydrodynamic spectrum together with a description of the modes (shear, diffusion, acoustic) in frequency space and obtain explicit values for the critical wave numbers.  

\section{Notation and Basic Definitions}\label{definitions}
Let $\mathcal{H}$ denote a Hilbert space and let $\mathbf{T}:\mathcal{H}\to\mathcal{H}$ be a closed linear operator with domain of definition $\mathcal{D}(\mathcal{H})$. We denote the spectrum of $\mathbf{T}$ as $\sigma(\mathbf{T})$ and its resolvent set as $\rho(\mathbf{T})$.\\
We denote the kernel of an operator $\mathbf{T}$ as $\ker(\mathbf{T})$ and its range as $\range(\mathbf{T})$. The codimension of a linear subspace $\mathcal{Y}\subseteq \mathcal{H}$ is defined as $\codim \mathcal{Y} = \dim (\mathcal{H})/\mathcal{Y}$, i.e., the dimension of its quotient space. A linear operator $\mathbf{T}$ is called \textit{semi-Fredholm} if $\range(\mathbf{T})$ is closed and $\dim\ker(\mathbf{T})$ or $\codim\range(\mathbf{T})$ are finite.\\
Following \cite{kato1995perturbation}, we define the \textit{essential spectrum} of $\mathbf{T}$ as 
\begin{equation}
\begin{split}
    \sigma_{\rm ess}(\mathbf{T}) & = \{\lambda \in \mathbb{C}: \range(\mathbf{T}-\lambda\mathbf{I}) \text{ is not closed or } \\
     &  \range(\mathbf{T}-\lambda\mathbf{I}) \text{ is closed, but } \dim\ker(\mathbf{T}-\lambda\mathbf{I})=\codim\range(\mathbf{T}-\lambda\mathbf{I})=\infty\},
    \end{split}
\end{equation}
where $\mathbf{I}$ denotes the identity operator on the Hilbert space $\mathcal{H}$, i.e., the essential spectrum of $\mathbf{T}$ consists of all complex numbers $\lambda$ such that $(\mathbf{T}-\lambda)$ is not semi-Fredholm.\\
The \textit{discrete spectrum} of an operator $\mathbf{T}$ is the set of its isolated eigenvalues of finite multiplicity,
\begin{equation}
    \sigma_{\rm disc}(\mathbf{T}) = \{\lambda\in\mathbb{C} : \ker(\mathbf{T}-\lambda\mathbf{I})\neq \emptyset,\quad \dim\ker(\mathbf{T}-\lambda\mathbf{I})<\infty, \quad \lambda \text{ isolated }  \}.
\end{equation}
We recall that the essential spectrum is stable under relatively compact perturbations \cite{hislop2012introduction} (Weyl's Theorem), i.e., $\sigma_{\rm ess}(\mathbf{T} + \mathbf{K}) = \sigma_{\rm ess}(\mathbf{T})$ for $\mathbf{K}$ compact. An operator $\mathbf{S}$ is called $\mathbf{T}$-degenerate if $\mathbf{S}$ is $\mathbf{T}$ - bounded and $\dim\range(\mathbf{S})$ is bounded.  \\


The spectral analysis of the main operator $\mathcal{L}$ of the paper (to be defined later) will be carried out on the Hilbert space 

\begin{equation}
\mathcal{H}_{\mathbb{T}^3,\mathbf{v}}=L^2_{\mathbf{x}}(\mathbb{T}^3) \times L^2_{\mathbf{v}}(\mathbb{R}^3,(2\pi)^{-\frac{3}{2}}e^{-\frac{|\mathbf{v}|^2}{2}}),
\end{equation}
together with the inner product 
\begin{equation}
\langle f, g \rangle_{\mathbb{T}^3,\mathbf{v}} = (2\pi)^{-\frac{3}{2}}\int_{\mathbb{T}^3}\int_{\mathbb{R}^3} f(\mathbf{x},\mathbf{v}) g^*(x,\mathbf{v})\,  e^{-\frac{|\mathbf{v}|^2}{2}} d\mathbf{v} d\mathbf{x},
\end{equation}
or on the Hilbert space
\begin{equation}
\mathcal{H}_{\mathbb{R}^3,\mathbf{v}}=L^2_{\mathbf{x}}(\mathbb{R}^3) \times L^2_{\mathbf{v}}(\mathbb{R}^3,(2\pi)^{-\frac{3}{2}}e^{-\frac{|\mathbf{v}|^2}{2}}),
\end{equation}
together with the inner product 
\begin{equation}
\langle f, g \rangle_{\mathbb{R}^3,\mathbf{v}} = (2\pi)^{-\frac{3}{2}}\int_{\mathbb{R}^3}\int_{\mathbb{R}^3} f(\mathbf{x},\mathbf{v}) g^*(x,\mathbf{v})\,  e^{-\frac{|\mathbf{v}|^2}{2}} d\mathbf{v} d\mathbf{x},
\end{equation}

where the star denotes complex conjugation. To treat the the torus and the whole space simultaneously in some estimates, we write $\langle f, g \rangle_{\mathbf{x},\mathbf{v}}$ to indicate either integration domain.\\
Because of the unitary properties of the Fourier transform and Fourier series, respectively, we can slice the space $\mathcal{H}$ for each wave number $\mathbf{k}$ and analyze the operator $\mathcal{L}_{\mathbf{k}}$ (restriction of $\mathcal{L}$ to the wave number $\mathbf{k}$) on the Hilbert space
\begin{equation}
\mathcal{H}_{\mathbf{v}} = L^2_{\mathbf{v}}(\mathbb{R}^3,(2\pi)^{-\frac{3}{2}}e^{-|\mathbf{v}|^2}),
\end{equation}
together with the inner product
\begin{equation}
\langle f, g \rangle_{\mathbf{v}} = (2\pi)^{-\frac{3}{2}}\int_{\mathbb{R}^3} f(\mathbf{v}) g^*(\mathbf{v}) e^{-\frac{|\mathbf{v}|^2}{2}} d\mathbf{v}.
\end{equation}
For a wave vector $\mathbf{k} \in \mathbb{R}^3$, we denote its wave number as
\begin{equation}
    k =  |\mathbf{k}|.
\end{equation}


\section{Preliminaries and Formulation of the Problem} \label{Prelim} 
\noindent
We will be concerned with the three-dimensional BGK kinetic equation
\begin{equation}\label{maineq}
\frac{\partial f}{\partial t}+\mathbf{v}\cdot\nabla_{\mathbf{x}} f=-\frac{1}{\tau}Q_{BGK},
\end{equation}
for the scalar distribution function either with the tours of width $L$ as spatial domain of definition, $f: \mathbb{T}^3_L\times\mathbb{R}^3\times [0,\infty)\to\mathbb{R}^{+}$, or the whole space, $f: \mathbb{R}^3\times\mathbb{R}^3\times [0,\infty)\to\mathbb{R}^{+}$, $f=f(\mathbf{x},\mathbf{v},t)$ and the BGK collision operator
\begin{equation}
Q_{BGK}=\Big(f(\mathbf{x},\mathbf{v},t)-f^{eq}(n[f],\mathbf{u}[f],T[f];\mathbf{v})\Big).
\end{equation}
Here, $\mathbb{T}^3_L$ denotes the three-dimensional torus of length $L$, the parameter $\tau>0$ is the relaxation time, the equilibrium distribution is given by the standard Gaussian
\begin{equation}
f^{eq}(n,\mathbf{u},T;\mathbf{v})=n\left(\frac{2\pi k_B T}{m}\right)^{-\frac{3}{2}}e^{-\frac{m}{2k_BT}|\mathbf{u}-\mathbf{v}|^2},
\end{equation}
for the molecular mass $m$ and the Boltzmann constant $k_B$, while the five scalar hydrodynamic variables are given by the number density,
\begin{equation}\label{defdensity}
n[f](\mathbf{x},t)=\int_{\mathbb{R}^3}f(\mathbf{x},\mathbf{v},t)\, d\mathbf{v},
\end{equation}
the velocity,
\begin{equation}\label{defvelocity}
\mathbf{u}[f](\mathbf{x},t)=\frac{1}{n[f](\mathbf{x},t)}\int_{\mathbb{R}^3}\mathbf{v}f(\mathbf{x},\mathbf{v},t)\, d\mathbf{v},
\end{equation}
and the temperature, which is defined implicitly through conservation of energy as
\begin{equation}\label{deftemperature}
\frac{3}{2}\frac{k_B}{m}T[f](\mathbf{x},t)n[f](\mathbf{x},t)+n[f](\mathbf{x},t)\frac{|\mathbf{u}[f](\mathbf{x},t)|^2}{2}=\int_{\mathbb{R}^3}\frac{|\mathbf{v}|^2}{2}f(\mathbf{x},\mathbf{v},t)\, d\mathbf{v}.
\end{equation}
The physical units are given as $[k_B]=m^2kg s^{-2}K^{-1}$ and $[k_B T] = m^2 kg s^{-2}$ respectively. 
We introduce the moments of the distribution function $f$ as
\begin{equation}\label{defmomentum}
\mathbf{M}^{(n)}(\mathbf{x},t)=\int_{\mathbb{R}^3}f(\mathbf{x},\mathbf{v},t)\,\mathbf{v}^{\otimes n}d\mathbf{v},
\end{equation}
where $\mathbf{v}^{\otimes 0}=1$, $\mathbf{v}^{\otimes 1}=\mathbf{v}$ and 
\begin{equation}
\mathbf{v}^{\otimes n}=\underbrace{\mathbf{v}\otimes...\otimes \mathbf{v}}_{n-\text{times}},
\end{equation}
for $n\geq 2$ is the $n$-th tensor power. The moment defined in \eqref{defmomentum} is an $n$-th order symmetric tensor, depending on space and time.\\
The first three moments relate to the hydrodynamic variables through
\begin{equation}\label{first3hydro}
\begin{split}
&\mathbf{M}^{(0)}=n,\\
&\mathbf{M}^{(1)}=n\mathbf{u},\\
&\tr\mathbf{M}^{(2)}=n\left(|\mathbf{u}|^2+3\frac{k_B T}{m}\right).
\end{split}
\end{equation}
Conversely, we can express the hydrodynamic variables in terms of the moments as
\begin{equation}
\begin{split}
n& = \mathbf{M}^{(0)},\\
\mathbf{u} &= \frac{\mathbf{M}_1}{\mathbf{M}^{(0)}},\\
 \frac{k_B}{m}T &= \frac{1}{3}\left(\frac{\tr \mathbf{M}_2}{\mathbf{M}^{(0)}}-\frac{|\mathbf{M}_1|^2}{(\mathbf{M}^{(0)})^2}\right).
\end{split}
\end{equation}
We can reformulate equation \eqref{maineq} as an infinite system of coupled momentum equations as\begin{equation}\label{momentumeq}\left(1+\tau\frac{\partial }{\partial t}\right)\mathbf{M}^{(n)}=-\tau\nabla\cdot\mathbf{M}^{(n+1)}+\mathbf{M}^{(n)}_{eq},\end{equation} for $n\geq 0$, where\begin{equation}\mathbf{M}^{(n)}_{eq}=\int_{\mathbb{R}^3}f^{eq}(n[f],\mathbf{u}[f],T[f];\mathbf{v})\mathbf{v}^{\otimes n}\,d\mathbf{v}.\end{equation}The special property of the BGK hierarchy is that the first three moment equations reduce to\begin{equation}\label{first3momenta}\begin{split}&\frac{\partial }{\partial t}\mathbf{M}^{(0)}=-\nabla\cdot\mathbf{M}^{(1)},\\&\frac{\partial }{\partial t}\mathbf{M}^{(1)}=-\nabla \cdot\mathbf{M}^{(2)},\\&\frac{\partial }{\partial t}\tr\mathbf{M}^{(2)}=-\tr(\nabla \cdot\mathbf{M}^{(3)}).\end{split}\end{equation}
In particular, the first three moment equations in terms of the hydrodynamic variables read
\begin{equation}\label{hydro}
\begin{split}
&\frac{\partial}{\partial t} n=-\nabla\cdot (n\mathbf{u}),\\
&\frac{\partial}{\partial t} (n\mathbf{u})=-\nabla\cdot \int_{\mathbb{R}^3}\mathbf{v}\otimes\mathbf{v} f\, d\mathbf{v},\\
&\frac{\partial}{\partial t}\left(\int_{\mathbb{R}^3}\frac{m|\mathbf{v}|^2}{2}f\, d\mathbf{v}\right)=-\nabla\cdot \int_{\mathbb{R}^3}\frac{|\mathbf{v}|^2}{2}\mathbf{v}f\, d\mathbf{v}.
\end{split}
\end{equation}

We will be interested in the linearized dynamics of \eqref{maineq} around a \textit{global} Maxwellian
\begin{equation}
\phi(\mathbf{v})= n_0 \left(2\pi \frac{k_BT_0}{m}\right)^{-\frac{3}{2}}e^{-\frac{m|\mathbf{v}|^2}{2k_BT_0}}.
\end{equation}
On the torus, we non-dimensionalize $\mathbf{x}\mapsto L\mathbf{x}$, which implies that $\mathbf{x}\in\mathbb{T}^3$ henceforth and introduce the \textit{thermal velocity} as well as the \textit{thermal time},
\begin{equation}\label{defvthermal}
v_{\rm thermal}=\sqrt{\frac{k_B T_0}{m}},\quad t_{\rm thermal} = L\sqrt{\frac{m}{k_BT_0}}
\end{equation}
for the equilibrium density $n_0$ and the equilibrium temperature $T_0$. On the whole space, there is no preferred macroscopic length scale and we may non-dimensionalize according to \eqref{defvthermal} with any reference length scale $L$. Re-scaling an non-dimensionalizing according to
\begin{equation}
\mathbf{v}\mapsto v_{\rm thermal}\mathbf{v},\qquad t\mapsto t t_{thermal},\qquad \tau \mapsto \tau t_{thermal},
\end{equation}
leads to the linearized, non-dimensional BGK equation
\begin{equation}\label{linBGK}
\frac{\partial f}{\partial t} = -\mathbf{v}\cdot \nabla_{\mathbf{x}}f-\frac{1}{\tau}f+\frac{1}{\tau}(2\pi)^{-3/2}e^{\frac{-|\mathbf{v}|^2}{2}}\left[\left(\frac{5}{2}-\frac{|\mathbf{v}|^2}{2}\right)M_0+\mathbf{M}_1\cdot\mathbf{v}+\frac{1}{6}(|\mathbf{v}|^2-3)\tr\mathbf{M}_2\right].
\end{equation}
Equation \eqref{linBGK} will be the starting point for further analysis. 
For later reference, we also define the mean free path as
\begin{equation}\label{defmeanfree}
l_{\rm mfp}=\tau v_{thermal}. 
\end{equation}

\section{Spectral Analysis of the linearized BGK operator}\label{spectralan}
In this section, we will carry out an explicit spectral analysis of the right-hand side of \eqref{linBGK}. This will allow us to draw conclusions on the decay properties of hydrodynamic variables, the existence of a critical wave number and the hydrodynamic closure. After reformulating the problem in frequency space, we will use the resolvent calculus to formulate a condition for the discrete spectrum (Subsection \ref{spectral1}). Then, we will use properties of the plasma dispersion function to define a spectral function $\Gamma_{\tau k}$, whose zeros coincide with the discrete, isolated eigenvalues (Subsection \ref{spectral2}). Then, in Subsection \ref{spectral3}, using Rouch\'{e}'s Theorem, we prove the existence of a critical wave number $k_{\rm crit}$ such that $\Gamma_{\tau k}$ has no zeros (i.e., there exists no eigenvalues) for $ k>k_{\rm crit}$. Finally, in Subsection \ref{spectral4}, we take a closer look at the branches of eigenvalues (modes) and their corresponding critical wave numbers. 

\subsection{The discrete spectrum of a finite-rank perturbation}\label{spectral1}

To ease notation, we define five distinguished vectors associated with the hydrodynamic moments as
\begin{equation}\label{base5}
\begin{split}
e_0(v) &= 1,\\
e_1(v) &=  v_1,\\
e_2(v) &= v_2,\\
e_3(v) &=  v_3,\\
e_4(v) &=  \frac{|\mathbf{v}|^2-3}{\sqrt{6}},
\end{split}
\end{equation} 
which satisfy the orthonormality condition,
\begin{equation}
\langle e_i, e_j \rangle_{\mathbf{v}} = \delta_{ij},\quad \text{ for } \quad  0 \leq i,j \leq 4,
\end{equation}
where $\delta_{ij}$ is the Kronecker's delta.
Defining
\begin{equation}\label{fcoef}
f_j=\langle e_j,f\rangle_{\mathbf{v}},
\end{equation}
we can define the following relations between the moments and the coefficients \eqref{fcoef}:
\begin{equation}
\begin{split}
\frac{5-|\mathbf{v}|^2}{2}M_0&=\frac{5-|\mathbf{v}|^2}{2}f_0=f_0e_0-\frac{\sqrt{6}}{2}f_0e_4,\\
\mathbf{v}\cdot \mathbf{M}_1 &=f_1e_1+f_2e_2+f_3e_3,\\
\frac{|\mathbf{v}|^2-3}{6}\tr\mathbf{M}_2&= e_4 \frac{1}{\sqrt{6}} \int_{\mathbb{R}} f |\mathbf{v}|^2\, d\mathbf{v}=e_4 \frac{1}{\sqrt{6}}\left( \int_{\mathbb{R}} f (|\mathbf{v}|^2-3)\, d\mathbf{v}+3M_0\right)\\
&=f_2e_4+\frac{3}{\sqrt{6}}f_0e_4.
\end{split}
\end{equation}
For compactness, we bundle these five basis polynomials into a single vector
\begin{equation}
\mathbf{e}=(e_0,e_1,e_2,e_3,e_4).
\end{equation}
To ease notation, we denote the projection onto the span of $\{e_j\}_{0\leq j \leq 4}$ as 
\begin{equation}
\mathbb{P}_5f = \sum_{j=0}^4 \langle f, e_j \rangle_{\mathbf{v}} e_j,
\end{equation}
for any $f\in \mathcal{H}_{\mathbf{v}}$. The linearized dynamics then takes the form
\begin{equation}\label{eqmainlinear}
\frac{\partial f}{\partial t} = \mathcal{L}f,
\end{equation}
for the linear operator
\begin{equation}\label{defL}
\mathcal{L} = -\mathbf{v}\cdot\nabla_{\mathbf{x}} - \frac{1}{\tau}+\frac{1}{\tau}\mathbb{P}_5.
\end{equation}
\begin{remark}
Let us recall that any function $f\in \mathcal{H}_{\mathbf{v}}$ admits a unique expansion as a multi-dimensional \textit{Hermite series}:
\begin{equation}\label{Hermite}
f(\mathbf{v})=\sum_{n=0}^{\infty} \mathbf{f}_n:\mathbf{H}_n(\mathbf{v}),
\end{equation}
where
\begin{equation}
\mathbf{H}_n = (-1)^n e^{\frac{|\mathbf{v}|^2}{2}}\nabla^{n}e^{\frac{-|\mathbf{v}|^2}{2}},
\end{equation}
\label{Check definition}
and $\mathbf{f}_n$ is an $n$-tensor. Since the five basis vectors \eqref{base5} appear in the expansion \eqref{Hermite} via an orthogonal splitting, we have that
\begin{equation}\label{ortho}
\langle \mathbb{P}_5f, (1-\mathbb{P}_5)f\rangle_{\mathbf{v}}=0,
\end{equation}
 for any $f\in\mathcal{H}_{\mathbf{v}}$. Hermite expansions were famously used by Grad in his seminal paper \cite{grad1949kinetic} to establish finite-moment closures.
\end{remark}
From 
\begin{equation}\label{dissipative}
\begin{split}
\langle \mathcal{L}f,f\rangle_{\mathbf{x},\mathbf{v}} & =\langle -\mathbf{v}\cdot\nabla_{\mathbf{x}}f - \frac{1}{\tau}f+\frac{1}{\tau}\mathbb{P}_5f,f\rangle_{\mathbf{x},\mathbf{v}}\\
&=\int_{\mathbb{D}^3}\int_{\mathbb{R}^3}(-\mathbf{v}\cdot\nabla_{\mathbf{x}}f - \frac{1}{\tau}f+\frac{1}{\tau}\mathbb{P}_{5}f) f e^{-\frac{|\mathbf{v}|^2}{2}}\,d\mathbf{x}d\mathbf{v}\\
&=\int_{\mathbb{D}^3}\int_{\mathbb{R}^3}-\frac{1}{\tau}[(1-\mathbb{P}_5)f] (\mathbb{P}_5f+(1-\mathbb{P}_5)f) e^{-\frac{|\mathbf{v}|^2}{2}}\,d\mathbf{x}d\mathbf{v}\\
&=-\frac{1}{\tau}\|(1-\mathbb{P}_5)f\|_{\mathbf{x},\mathbf{v}}^2,
\end{split}
\end{equation}
where $\mathbb{D}^3\in \{\mathbb{T}^3,\mathbb{R}^3\}$ and we have assumed that $f$ is sufficiently regular to justify the application of the divergence theorem in $\mathbf{x}$ in order to remove the gradient term as well as \eqref{ortho}, it follows that the operator $\mathcal{L}$ is dissipative and that 
\begin{equation}
\Re\sigma(\mathcal{L})\leq 0.
\end{equation}
On the other hand, from \eqref{dissipative} and from $\|1-\mathbb{P}_5\|_{op}=1$, since $1-\mathbb{P}_5$ is a projection as well, it follows that
\begin{equation}\label{lowerbound}
    \langle \mathcal{L}f,f\rangle_{\mathbf{x},\mathbf{v}} \geq -\frac{1}{\tau} \|f\|_{\mathbf{x},\mathbf{v}}^2.
\end{equation}
This shows that any solution to \eqref{eqmainlinear} has to converge to zero, i.e., the global Maxwellian is a stable equilibrium up to the conserved quantities from the center mode.  On the other hand, we infer that the overall convergence rate to equilibrium can be at most $-\frac{1}{\tau}$, which immediately implies that there cannot be any eigenvalues below the essential spectrum (see also the next section).\\

Let us proceed with the spectral analysis by switching to frequency space. For $\mathbf{x}\in\mathbb{T}^3$, we can decompose $f$ in a Fourier series as
\begin{equation}
f(\mathbf{x},\mathbf{v})= \sum_{ k=0}^{\infty}\hat{f}(\mathbf{k},\mathbf{v}) e^{\ri \mathbf{x}\cdot\mathbf{k}}, 
\end{equation}
for the Fourier coefficients
\begin{equation}
\hat{f}(\mathbf{k},\mathbf{v}) = \frac{1}{(2\pi)^3}\int_{\mathbb{R}^3} f(\mathbf{x},\mathbf{v})e^{-\ri \mathbf{x}\cdot\mathbf{k}}\, d\mathbf{x},
\end{equation}
while on the whole space, we write
\begin{equation}
    f(\mathbf{x},\mathbf{v}) =  \frac{1}{(2\pi)^3}\int_{\mathbb{R}^3} \hat{f}(\mathbf{k},\mathbf{v})e^{\ri \mathbf{x}\cdot\mathbf{k}}\, d\mathbf{k},
\end{equation}
for the Fourier transform
\begin{equation}
    \hat{f}(\mathbf{k},\mathbf{v}) = \int_{\mathbb{R}^3} f(\mathbf{x},\mathbf{v})e^{-\ri \mathbf{x}\cdot\mathbf{k}}\, d\mathbf{x},
\end{equation}

In frequency space, for either the torus or the whole space, the operator \eqref{defL} is conjugated to the linear operator
\begin{equation}
 \mathcal{L}_{\mathbf{k}} f = -\ri (\mathbf{v}\cdot \mathbf{k})f-\frac{1}{\tau}f+\frac{1}{\tau}\mathbb{P}_5f,
\end{equation}
which implies that
\begin{equation}
\sigma (\mathcal{L}) =  \bigcup_{\mathbf{k}\in \mathbb{Z}^3} \sigma( \mathcal{L}_{\mathbf{k}}) \quad \text{or} \quad \bigcup_{\mathbf{k}\in \mathbb{R}^3} \sigma( \mathcal{L}_{\mathbf{k}}).  
\end{equation}

First, let us take a look at the spectrum of $ \mathcal{L}_0$. For $\mathbf{k}=0$, we see that $ \mathcal{L}$ collapses to a diagonal operator with five dimensional kernel spanned by $\{e_j\}_{0\leq j\leq 4}$:
\begin{equation}
 \mathcal{L}_0e_j=-\frac{1}{\tau}(e_j-\mathbb{P}_5e_j)=0,\quad 0\leq j\leq 4.
\end{equation}
On the other hand, the operator $ \mathcal{L}_0$ acts just like $-\frac{1}{\tau}$ on the orthogonal complement of $\text{span}\{e_j\}_{0\leq j\leq 4}$. This shows that 
\begin{equation}\label{sigma0}
\sigma( \mathcal{L}_0)=\left\{-\frac{1}{\tau},0\right\},
\end{equation}
where the eigenspace associated to zero has dimension five, while the eigenspace associated to $-\frac{1}{\tau}$ has co-dimension five. Furthermore, since $\range(\mathcal{L}_0) = \{f : \mathbb{P}_5 f = 0\}$ we find that $\codim\, \range\, \mathcal{L}_0 = 5$ and hence $\ind\, \mathcal{L}_0 = 0$, while $\dim\range (\mathcal{L}_0+\frac{1}{\tau}) = 5$ thus proving that $ (\mathcal{L}_0+\frac{1}{\tau})$ is not Fredholm, which implies the splitting 
\begin{equation}
    \sigma_{\rm disc}(\mathcal{L}_0)=\{0\},\quad \sigma_{\rm ess}(\mathcal{L}_0) = \left\{-\frac{1}{\tau}\right\}.
\end{equation}
Now, let us analyse $\mathcal{L}_{\mathbf{k}}$ for $\mathbf{k}\neq 0$. To ease notation in the following argument, we define the operator
\begin{equation}
\mathcal{S}_{\mathbf{k}}f = \mathbf{v}\cdot\mathbf{k} f,
\end{equation}
for any $\mathbf{k}\neq 0$, which gives
\begin{equation}
\begin{split}
\sigma( \mathcal{L}_{\mathbf{k}})&= -\frac{1}{\tau} - \sigma\left(\ri \mathcal{S}_k-\frac{1}{\tau}\mathbb{P}_5\right)\\
&=-\frac{1}{\tau} - \frac{1}{\tau}\sigma\left(\ri \tau\mathcal{S}_k-\mathbb{P}_5\right).
\end{split}
\end{equation}
Because the resolvent of $\mathcal{S}_{\mathbf{k}}$ is just given by multiplication with $(\mathbf{v}\cdot\mathbf{k}-z)^{-1}$, we see immediately that $\sigma(\mathcal{S}_{\mathbf{k}})=\mathbb{R}$, see also \cite{teschl2000jacobi}. We define the Green's function matrices as
\begin{equation}
\begin{split}
G_T(z,n,m) &= \langle (\ri \tau \mathcal{S}_{\mathbf{k}}-\mathbb{P}_5-z)^{-1}e_n,e_m\rangle_{\mathbf{v}},\\
G_S(z,n,m) &= \langle (\ri\tau\mathcal{S}_{\mathbf{k}}-z)^{-1}e_n,e_m\rangle_{\mathbf{v}},
\end{split}
\end{equation}
for $0\leq n,m \leq 4$ and set $G_S(z)=\{G_S(z,n,m)\}_{0\leq n,m\leq 4}$, $G_T(z)=\{G_T(z,n,m)\}_{0\leq n,m\leq 4}$.\\
By the second resolvent identity,
\begin{equation}
\mathcal{R}(z;A)-\mathcal{R}(z;B)=\mathcal{R}(z;A)(B-A)\mathcal{R}(z;B),
\end{equation}
for any operators $A,B$ and $z\in\rho(A)\cap\rho(B)$, we have for $A=\ri\tau\mathcal{S}_{\mathbf{k}}$ and $B=\ri\tau\mathcal{S}_{\mathbf{k}}-\mathbb{P}_5$ that
\begin{equation}\label{res1}
(\ri\tau\mathcal{S}_{\mathbf{k}}-\mathbb{P}_5-z)^{-1}=(\ri\tau\mathcal{S}_{\mathbf{k}}-z)^{-1}+(\ri\tau\mathcal{S}_{\mathbf{k}}-z)^{-1}\mathbb{P}_5(\ri\tau\mathcal{S}_{\mathbf{k}}-\mathbb{P}_5-z)^{-1}.
\end{equation}
Applying equation \eqref{res1} to $e_m$ for $0\leq m\leq 4$ and rearranging gives
\begin{equation}\label{resdelta}
\begin{split}
(\ri\tau\mathcal{S}_{\mathbf{k}}-\mathbb{P}_5-z)^{-1}e_n&=(\ri\tau\mathcal{S}_{\mathbf{k}}-z)^{-1}e_n+ (\ri\tau\mathcal{S}_{\mathbf{k}}-z)^{-1}\mathbb{P}_5(\ri\tau\mathcal{S}_{\mathbf{k}}-\mathbb{P}_5-z)^{-1}e_n\\
&=(\ri\tau\mathcal{S}_{\mathbf{k}}-z)^{-1}e_n+(\ri\tau\mathcal{S}_{\mathbf{k}}-z)^{-1}\sum_{j=0}^4\langle(\ri\tau\mathcal{S}_{\mathbf{k}}-\mathbb{P}_5-z)^{-1}e_n,e_j\rangle_{\mathbf{v}}e_j\\
&=(\ri\tau\mathcal{S}_{\mathbf{k}}-z)^{-1}e_n+\sum_{j=0}^4G_T(z,n,j)(\ri\tau\mathcal{S}_{\mathbf{k}}-z)^{-1}e_j,
\end{split}
\end{equation}
for $z\in\mathbb{C}\setminus\ri\mathbb{R}$. Thus, the resolvent of $\ri\tau\mathcal{S}_{\mathbf{k}}-\mathbb{P}_5-z$ includes the resolvent of $\ri\tau\mathcal{S}_{\mathbf{k}}$ as well as information from the matrix $\{G_T(z,n,m)\}_{0\leq n,m\leq 4}$ as coefficients.\\
Taking an inner product of \eqref{resdelta} with $e_m$ gives
\begin{equation}\label{eqGSGT}
\begin{split}
G_T(z,n,m)&=G_S(z,n,m)+\sum_{j=0}^4G_T(z,n,j)\langle (\ri\tau\mathcal{S}_{\mathbf{k}}-z)^{-1}e_j ,e_m\rangle_{\mathbf{v}}\\
&=G_S(z,n,m)+\sum_{j=0}^4G_T(z,n,j)G_S(z,j,m)
\end{split}
\end{equation}
for $0\leq n,m\leq 4$ and $z\in\mathbb{C}\setminus\ri\mathbb{R}$. System \eqref{eqGSGT} defines twenty-five equations for the coefficients $G_T(z,n,m)$, which can be re-written more compactly as
\begin{equation}
G_T=G_S+G_TG_S,
\end{equation}
or, equivalently,
\begin{equation}\label{IdG_S}
G_T(\Id-G_S)=G_S.
\end{equation}
Equation \eqref{IdG_S} can be interpreted as a special case of Krein's resolvent identity \cite{kurasov2004krein}. This shows that we can solve for the entries of $G_T$ unless $\det(\Id-G_S)=0$, or, to phrase it differently, we have that for each wave number $\mathbf{k}$, the discrete spectrum of $(\ri\tau\mathcal{S}_{\mathbf{k}})-\mathbb{P}_5$ can be used to infer that
\begin{equation}\label{spec1}
\sigma_{\rm disc}( \mathcal{L}_{\mathbf{k}})=-\frac{1}{\tau}-\frac{1}{\tau}\left\{z\in\mathbb{C}:\det\left(\int_{\mathbb{R}^3}\mathbf{e}(\mathbf{v})\otimes\mathbf{e}(\mathbf{v}) \frac{e^{-\frac{|\mathbf{v}|^2}{2}}}{\ri\tau \mathbf{k}\cdot\mathbf{v}-z}\, d\mathbf{v}-\Id\right)=0\right\}.
\end{equation}

An eigenvalue $\lambda$ of the operator $ \mathcal{L}_{\mathbf{k}}$ is related to the zero $z$ in \eqref{spec1} via
\begin{equation}\label{lambdaz}
z=-\tau\lambda-1.
\end{equation}
In particular, the finite-rank perturbation $\mathbb{P}_5$ can only add discrete eigenvalues to the spectrum and we have that $\sigma_{ess}(\ri\tau\mathcal{S}_{\mathbf{k}}-\mathbb{P}_5)=\sigma_{ess}(\ri\tau\mathcal{S}_{\mathbf{k}})=\ri\mathbb{R}$. \\

\begin{remark}
As discussed in \cite{kato1995perturbation}, the determinant appearing in \eqref{spec1} is called \textit{Weinstein--Aronszajn determinant} for the absolutely degenerate perturbation $\mathbb{P}_5$. The determinant as function $z$ is meromorphic function on the resolvent set of the multiplication operator $\ri(\mathbf{v}\cdot\mathbf{k})$. Its zeros correspond to isolated eigenvalues of $\mathcal{L}_{\mathbf{k}}$ and the multiplicity as eigenvalue equals the order of the zero of the determinant, see \cite{kato1995perturbation}. 
\end{remark}

\subsection{Reformulation in terms of the spectral function}\label{spectral2}
We proceed with the spectral analysis of \eqref{defL} by rewriting the determinant expression in \eqref{spec1}. To this end, we note that any wave vector $\mathbf{k}\in\mathbb{R}^3$ can be written as
\begin{equation}
\mathbf{k}=\mathbf{Q}_{\mathbf{k}}( k,0,0)^T,
\end{equation}
for a unique rotation matrix $\mathbf{Q}_{\mathbf{k}}$. Defining $\mathbf{w}=\mathbf{Q}_{\mathbf{k}}^T\mathbf{v}$, we have that
\begin{equation}
\mathbf{k}\cdot\mathbf{v}=\mathbf{Q}_{\mathbf{k}}( k,0,0)^T\cdot\mathbf{v} = ( k,0,0)\cdot\mathbf{w} =  kw_1,
\end{equation}
while the vector of basis functions $\mathbf{e}$ transforms according to 
\begin{equation}
\begin{split}
\mathbf{e}(\mathbf{v})&=(2\pi)^{-\frac{3}{4}}\left(1,\mathbf{v},\frac{|\mathbf{v}|^2-3}{\sqrt{6}}\right)=(2\pi)^{-\frac{3}{4}}\left(1,\mathbf{Q}_{\mathbf{k}}\mathbf{w},\frac{|\mathbf{w}|^2-3}{\sqrt{6}}\right)\\
&=
\begin{pmatrix}
1 & 0 & 0\\
0 & \mathbf{Q}_{\mathbf{k}} & 0\\
0 & 0 & 1
\end{pmatrix}\mathbf{e}(\mathbf{w}).
\end{split}
\end{equation}
This, together with $d\mathbf{v}=d\mathbf{w}$ from the orthogonality of $\mathbf{Q}_{\mathbf{k}}$, implies that
\begin{equation}
\begin{split}
&\det\left(\int_{\mathbb{R}^3}\mathbf{e}(\mathbf{v})\otimes\mathbf{e}(\mathbf{v}) \frac{e^{-\frac{|\mathbf{v}|^2}{2}}}{\ri\tau \mathbf{k}\cdot\mathbf{v}-z}\, d\mathbf{v}-\Id\right)\\
&\qquad=\det\left(\int_{\mathbb{R}^3}\begin{pmatrix}
1 & 0 & 0\\
0 & \mathbf{Q}_{\mathbf{k}} & 0\\
0 & 0 & 1
\end{pmatrix}\mathbf{e}(\mathbf{w})\otimes\left(\begin{pmatrix}
1 & 0 & 0\\
0 & \mathbf{Q}_{\mathbf{k}} & 0\\
0 & 0 & 1
\end{pmatrix}\mathbf{e}(\mathbf{w})\right)\frac{e^{-\frac{|\mathbf{w}|^2}{2}}}{\ri\tau  kw_1-z}\, d\mathbf{w}-\Id\right)\\
&\qquad=\det\left(\int_{\mathbb{R}^3}\mathbf{e}(\mathbf{w})\otimes\mathbf{e}(\mathbf{w}) \frac{e^{-\frac{|\mathbf{w}|^2}{2}}}{\ri\tau  kw_1-z}\, d\mathbf{w}-\Id\right),\\
\end{split}
\end{equation}
where we have used the orthogonality of $\mathbf{Q}_{\mathbf{k}}$.\\

We proceed:
\begin{equation}\label{det1}
\begin{split}
&\det\left(\int_{\mathbb{R}^3}\mathbf{e}(\mathbf{w})\otimes\mathbf{e}(\mathbf{w}) \frac{e^{-\frac{|\mathbf{w}|^2}{2}}}{\ri\tau  kw_1-z}\, d\mathbf{w}-\Id\right)=\\[0.33cm]
&\qquad  =\det\left[(2\pi)^{-\frac{3}{2}}\int_{\mathbb{R}^3} \begin{pmatrix}
1 & w_1 & w_2 & w_3 & \frac{|\mathbf{w}|^2-3}{\sqrt{6}}\\
w_1 & w_1^2 & w_1w_2 & w_1w_3 & w_1 \frac{|\mathbf{w}|^2-3}{\sqrt{6}}\\
w_2 & w_1w_2 & w_2^2 & w_2w_3 & w_2 \frac{|\mathbf{w}|^2-3}{\sqrt{6}}\\
w_3 & w_1w_3 & w_3w_2 & w_3^2 & w_3 \frac{|\mathbf{w}|^2-3}{\sqrt{6}}\\
\frac{|\mathbf{w}|^2-3}{\sqrt{6}} & w_1\frac{|\mathbf{w}|^2-3}{\sqrt{6}} & w_2\frac{|\mathbf{w}|^2-3}{\sqrt{6}} & w_3\frac{|\mathbf{w}|^2-3}{\sqrt{6}} & \frac{(|\mathbf{w}|^2-3)^2}{6}
\end{pmatrix}
\frac{e^{-\frac{|\mathbf{w}|^2}{2}}}{\ri\tau  kw_1-z}\, d\mathbf{w}-\Id\right].
\end{split}
\end{equation}
Integrating out the variables $w_2$ and $w_3$, it follows,
\begin{equation}\label{det2}
\begin{split}
&\det\left(\int_{\mathbb{R}^3}\mathbf{e}(\mathbf{w})\otimes\mathbf{e}(\mathbf{w}) \frac{e^{-\frac{|\mathbf{w}|^2}{2}}}{\ri\tau  kw_1-z}\, d\mathbf{w}-\Id\right)=\\[0.33cm]
&\qquad  =\det\left[(2\pi)^{-\frac{3}{2}}\int_{\mathbb{R}} \begin{pmatrix}
2\pi & 2\pi w_1 & 0 & 0 & 2\pi\frac{w_1^2-1}{\sqrt{6}}\\
2\pi w_1 & 2\pi w_1^2 & 0 & 0 & 2\pi w_1 \frac{w_1^2-1}{\sqrt{6}}\\
0 & 0 & 2\pi & 0 & 0\\
0 & 0 & 0 & 2\pi & 0\\
2\pi\frac{w_1^2-1}{\sqrt{6}} & 2\pi w_1\frac{w_1^2-1}{\sqrt{6}} & 0 & 0 & 2\pi \frac{w_1^4-2w_1^2+5}{6}
\end{pmatrix}
\frac{e^{-\frac{w_1^2}{2}}}{\ri\tau  kw_1-z}\, dw_1-\Id\right]\\[0.33cm]
&\qquad  =\det\left[\frac{1}{\sqrt{2\pi}}\int_{\mathbb{R}}\begin{pmatrix}
1 & w &  \frac{w^2-1}{\sqrt{6}}\\
w& w^2 &   w \frac{w^2-1}{\sqrt{6}}\\
\frac{w^2-1}{\sqrt{6}} & w\frac{w^2-1}{\sqrt{6}} & \frac{w^4-2w^2+5}{6}
\end{pmatrix}
\frac{e^{-\frac{w^2}{2}}}{\ri\tau  kw-z}\, dw-\Id\right]\left(\frac{1}{\sqrt{2\pi}}\int_{\mathbb{R}}\frac{e^{-\frac{w^2}{2}}}{\ri\tau k w-z}{dw}-1\right)^2,
\end{split}
\end{equation}
where we have used the linearity of the integral and properties of the determinant of block matrices and changed the notation of the integration variable $w_1\mapsto w$ to ease notation. 
Also, we have used that
\begin{equation}
\begin{split}
\int_{\mathbb{R}^2} &(w_1^2+w_2^2+w_3^2-3)^2 e^{-\frac{w_2^2}{2}-\frac{w_3^2}{2}}\, dw_2dw_3\\
&\qquad= \int_{\mathbb{R}^2} (w_1^4+w_2^4+w_3^4+9 -6w_1^2-6w_2^2-6w_3^2 +2w_1^2w_2^2+2w_2^2w_3^2+2w_1^2w_3^2)e^{-\frac{w_2^2}{2}-\frac{w_3^2}{2}}\, dw_2dw_3\\
&\qquad = 2\pi\Big(w_1^4 + 3 +3 + 9 - 6w_1^2 -6 -6 +2w_1^2 +2 + 2w_1^2 \Big)\\
&\qquad = 2\pi\Big(w_1^4 -2w_1^2 +5 \Big).
\end{split}
\end{equation}

For the following calculation, let us define the function
\begin{equation}\label{I0}
Z(\zeta)=\frac{1}{\sqrt{2\pi}}\int_{\mathbb{R}}\frac{e^{-\frac{v^2}{2}}}{v-\zeta}\, dv ,
\end{equation}
for $\zeta\in \mathbb{C}\setminus\mathbb{R}$. From \eqref{dissipative}, it suffices to consider $Z$ for $\Im \zeta>0$. The symmetry properties
\begin{equation}\label{symZ}
    Z(\zeta^*)=Z^*(\zeta),\quad  Z(-\zeta)=-Z(\zeta),
\end{equation}
which can be inferred directly from \eqref{I0}, however, allow us to extend the function to the whole complex plane (with a discontinuity at the real line) once an expression for a half-plane is known.

\begin{remark}
Integral expressions of the form \eqref{I0} appear frequently in thermodynamics and plasma physics \cite{fitzpatrick2014plasma}, where the function \eqref{I0} is called \textit{plasma dispersion function} accordingly. It admits an explicit representation as
\begin{equation}\label{g0}
Z(\zeta)=\ri\sqrt{\frac{\pi}{2}} e^{-\frac{\zeta^2}{2}}\left[\sign(\Im{\zeta})-\erf\left(\frac{-\ri \zeta}{\sqrt{2}}\right)\right], \quad \Im{\zeta}\neq 0,
\end{equation}
see, e.g., \cite{abramowitz1948handbook} and solves the differential equation
\begin{equation}\label{dI0}
\frac{d}{d\zeta}Z= -\zeta Z-1.
\end{equation}
In the following, we will be concerned mostly with $Z$ on the upper half-plane, thus defining
\begin{equation}\label{defIplus}
    Z_{+}(\zeta)=\ri\sqrt{\frac{\pi}{2}} e^{-\frac{\zeta^2}{2}}\left[1-\erf\left(\frac{-\ri \zeta}{\sqrt{2}}\right)\right], \quad \Im{\zeta}> 0. 
\end{equation}
Using a generalization of Watson's Lemma, one can prove the asymptotic expansion 
 \begin{equation}\label{Ipsymptotic}
 Z_+(\zeta) \sim -\sum_{n=0}^\infty \frac{(2n-1)!!}{\zeta^{2n+1}},  \qquad \text{ for }|\arg(\zeta)|\leq \frac{\pi}{2}-\delta,\qquad  \zeta\to\infty ,
 \end{equation}
for any  $0<\delta\leq \frac{\pi}{2}$ and $\Im\zeta>0$, see \cite{huba1998nrl}.

\end{remark}

Using the recurrence relation \eqref{dI0}, we calculate the first few derivatives of $Z$ in terms of polynomials and $Z$ itself:
\begin{equation}\label{diffI0}
\begin{split}
\frac{dZ}{d\zeta}&=-1-\zeta Z,\\
\frac{d^2Z}{d\zeta^2}&=\zeta+(\zeta^2-1)Z,\\
\frac{d^3Z}{d\zeta^3}&=2-\zeta^2+(3\zeta-\zeta^2)Z,\\
\frac{d^4Z}{d\zeta^4}&=-5\zeta+\zeta^3+(\zeta^4-6\zeta^2+3)Z.
\end{split}
\end{equation}
Using the identity
\begin{equation}\label{HermiteInt}
\begin{split}
\frac{1}{\sqrt{2\pi}}\int_{\mathbb{R}} H_k(v)\frac{e^{-\frac{v^2}{2}}}{v-z}\, dv& = \frac{1}{\sqrt{2\pi}}\int_{\mathbb{R}}\left[\left(-\frac{d}{dv}\right)^ke^{-\frac{v^2}{2}}\right]\frac{dv}{v-z}= \frac{(-1)^kk!}{\sqrt{2\pi}}\int_{\mathbb{R}}e^{-\frac{v^2}{2}}\frac{dv}{(v-z)^{k+1}}\\
&=\frac{(-1)^k}{\sqrt{2\pi}}\frac{d^k}{dz^k}\int_{\mathbb{R}}e^{-\frac{v^2}{2}}\frac{dv}{v-z}= (-1)^k\frac{d^kZ}{dz^k},
\end{split}
\end{equation}
together with \eqref{diffI0} allows us to further simplify the determinant expression in \eqref{det2}. Indeed, expanding the polynomial matrix in \eqref{det1} in Hermite basis and using \eqref{HermiteInt}, we define the matrix,
\begin{equation}\label{expandmatrix}
\begin{split}
M(\zeta) &:=
\frac{1}{\sqrt{2\pi}}\int_{\mathbb{R}}\begin{pmatrix}
1 & w &  \frac{w^2-1}{\sqrt{6}}\\
w& w^2 &   w \frac{w^2-1}{\sqrt{6}}\\
\frac{w^2-1}{\sqrt{6}} & w\frac{w^2-1}{\sqrt{6}} & \frac{w^4-2w^2+5}{6}
\end{pmatrix}\frac{e^{-\frac{w^2}{2}}}{w-\zeta}\, dw\\
&=\frac{1}{\sqrt{2\pi}}
\int_{\mathbb{R}} \begin{pmatrix}
H_0(w) & H_1(w) &  \frac{H_2(w)}{\sqrt{6}}\\
H_1(w)& H_2(w)+H_0(w) &   \frac{H_3(w)+2H_1(w)}{\sqrt{6}}\\
\frac{H_2(w)}{\sqrt{6}} & \frac{H_3(w)+2H_1(w)}{\sqrt{6}} & \frac{H_4(w)+4H_2(w)+6}{6}
\end{pmatrix} \frac{e^{-\frac{w^2}{2}}}{w-\zeta}\, dw\\
&=\begin{pmatrix}
Z & -Z' &  \frac{Z''}{\sqrt{6}}\\
-Z' & Z''+Z &   -\frac{Z'''+2Z'}{\sqrt{6}}\\
\frac{Z''}{\sqrt{6}} & -\frac{Z'''+Z'}{\sqrt{6}} & \frac{Z^{(4)}+4Z''+6H_0}{6}
\end{pmatrix}\\
&=\begin{pmatrix}
Z & 1+\zeta Z &  \frac{\zeta +(\zeta^2-1)Z}{\sqrt{6}}\\
1+\zeta Z & \zeta + \zeta^2 Z &   \frac{\zeta^2+(\zeta^3-\zeta)Z}{\sqrt{6}}\\
\frac{\zeta +(\zeta^2-1)Z}{\sqrt{6}} & \frac{\zeta^2+(\zeta^3-\zeta)Z}{\sqrt{6}}& \frac{\zeta^3-\zeta+(\zeta^4-2\zeta^2+5)Z}{6}
\end{pmatrix}.
\end{split}
\end{equation}
\label{insert cubic scaling in k}
Let $I$ denote the $3\times 3$ identity matrix. To ease notation, we define the \textit{spectral function},
\begin{equation}\label{defchar}
\begin{split}
\Gamma_{\tau k}(\zeta)&:= \det(M(\zeta)-\ri\tau k I)\\
&=\det\begin{pmatrix}
Z(\zeta)-\ri\tau  k & 1+\zeta Z(\zeta) &  \frac{\zeta +(\zeta^2-1)Z(\zeta)}{\sqrt{6}}\\
1+\zeta Z(\zeta)  & \zeta+\zeta^2 Z(\zeta)-\ri\tau  k &   \frac{\zeta^2+(\zeta^3-\zeta)Z(\zeta)}{\sqrt{6}}\\
\frac{\zeta +(\zeta^2-1)Z(\zeta)}{\sqrt{6}} & \frac{\zeta^2+(\zeta^3-\zeta)Z(\zeta)}{\sqrt{6}}& \frac{\zeta^3-\zeta+(\zeta^4-2\zeta^2+5)Z(\zeta)}{6}-\ri\tau  k
\end{pmatrix}\\
&=\frac{1}{6}\left(\zeta+6 \ri  k^3 \tau ^3-\zeta  (\zeta^2+5)  k^2 \tau ^2+2 \ri (\zeta ^2+3)  k \tau \right.\\
&\qquad\left.+Z(\zeta ) (\zeta ^2-(\zeta^4+4 \zeta ^2+11)  k^2 \tau ^2+2 \ri \zeta ^3  k \tau -5)-4 \ri Z^2 (\zeta )((\zeta ^2+1)  k \tau -\ri \zeta ) \right),
\end{split}
\end{equation}
which allows us to conclude that
\begin{equation}\label{detexpr}
\det\left(\int_{\mathbb{R}^3}\mathbf{e}(\mathbf{w})\otimes\mathbf{e}(\mathbf{w}) \frac{e^{-\frac{|\mathbf{v}|^2}{2}}}{\ri\tau\mathbf{k}\cdot \mathbf{v}-z}\, d\mathbf{v}-\Id\right)
=\frac{1}{(\ri k\tau)^5}(Z(\zeta)-\ri\tau k)^2\Gamma_{\tau k}(\zeta)\bigg|_{\zeta=\frac{z}{\ri k\tau}},
\end{equation}
by the scaling properties of the determinant function. Consequently, from \eqref{spec1} and \eqref{lambdaz} we deduce that,
\begin{equation}\label{spec2}
\begin{split}
\sigma_{\rm disc}( \mathcal{L}_{\mathbf{k}}) = \left\{\lambda\in\mathbb{C}:\Gamma_{\tau k}\left(\frac{-\tau\lambda-1}{\ri k\tau}\right)=0
\right\} \cup \left\{\lambda\in\mathbb{C}:Z\left(\frac{-\tau\lambda-1}{\ri k\tau}\right)=\ri\tau k \right\}.
\end{split}
\end{equation}
Typical spectra \eqref{spec2}
for different wave numbers are shown in Figures \ref{plotspec1} - \ref{plotspec3}. 
\begin{remark}
    Let us emphasize that the formulas for the discrete spectrum, as derived from the Weinstein--Aronszajn determinant and the plasma dispersion function in \eqref{spec2} are well-defined in the limit $k\to 0$. Indeed, the thrice degenerate eigenvalue $\lambda = 0$ corresponds so $\zeta = \ri/(k\tau)$ in the upper half-plane from which the four eigenvalue branches bifurcate. By Taylor expanding the spectral function \eqref{defchar} in $k$, we obtain the leading order contributions of the eigenvalues for small $k$:
    \begin{equation}
      \begin{split}
    \lambda_{\rm diff}(k) & = -\tau k^2+\mathcal{O}(k^4),\\
    \lambda_{\rm shear}(k) & = -\tau k^2+\mathcal{O}(k^4),\\
    \lambda_{\rm ac}(k) & = \ri\sqrt{\frac{5}{3}}k -\tau k^2 +\mathcal{O}(k^3),
    \end{split}
    \end{equation}
    which are consistent with Chapman--Enskog expansion, see \cite{cercignani1988boltzmann}. 
\end{remark}
\begin{remark}\label{remsym}
   The symmetry properties of the plasma dispersion function \eqref{symZ} together with the structure of \eqref{defchar} allows us to infer a the symmetry property
   \begin{equation}\label{symGam}
       \Gamma_{\tau k}(-\zeta^{*}) = -\Gamma_{\tau k}(\zeta)^{*},
   \end{equation}
   which implies that if $\zeta$ is a zero to $\Gamma_{\tau k}$, so is $-\zeta^{*}$. Consequently, since
   \begin{equation}
       \zeta = -\frac{\tau\lambda+1}{\ri k \tau},
   \end{equation}
   we find that if $\lambda$ is an eigenvalue, so is $\lambda^*$. In particular, the eigenvalues are real or come in complex conjugated pairs for each wave number $k$.  
\end{remark}
The explicit transcendental equation \eqref{spec2} determining the discrete spectrum is the first main result of our paper. It will allow us to draw further conclusions about the discrete (hydrodynamic) spectrum.

\subsection{Existence of a Critical Wave Number and Finiteness of the Hydrodynamic Spectrum}\label{spectral3}
Next, let us prove that there exists a critical wave number ${k}_{\rm crit}$, such that
\begin{equation}
    \sigma_{\rm disc}( \mathcal{L}_{\mathbf{k}})=\emptyset,\quad \text{ for }  k> k_{\rm crit}. 
\end{equation}

\begin{centering}
\begin{figure}
	\begin{subfigure}{.5\textwidth}
		\centering
		\includegraphics[width=1\linewidth]{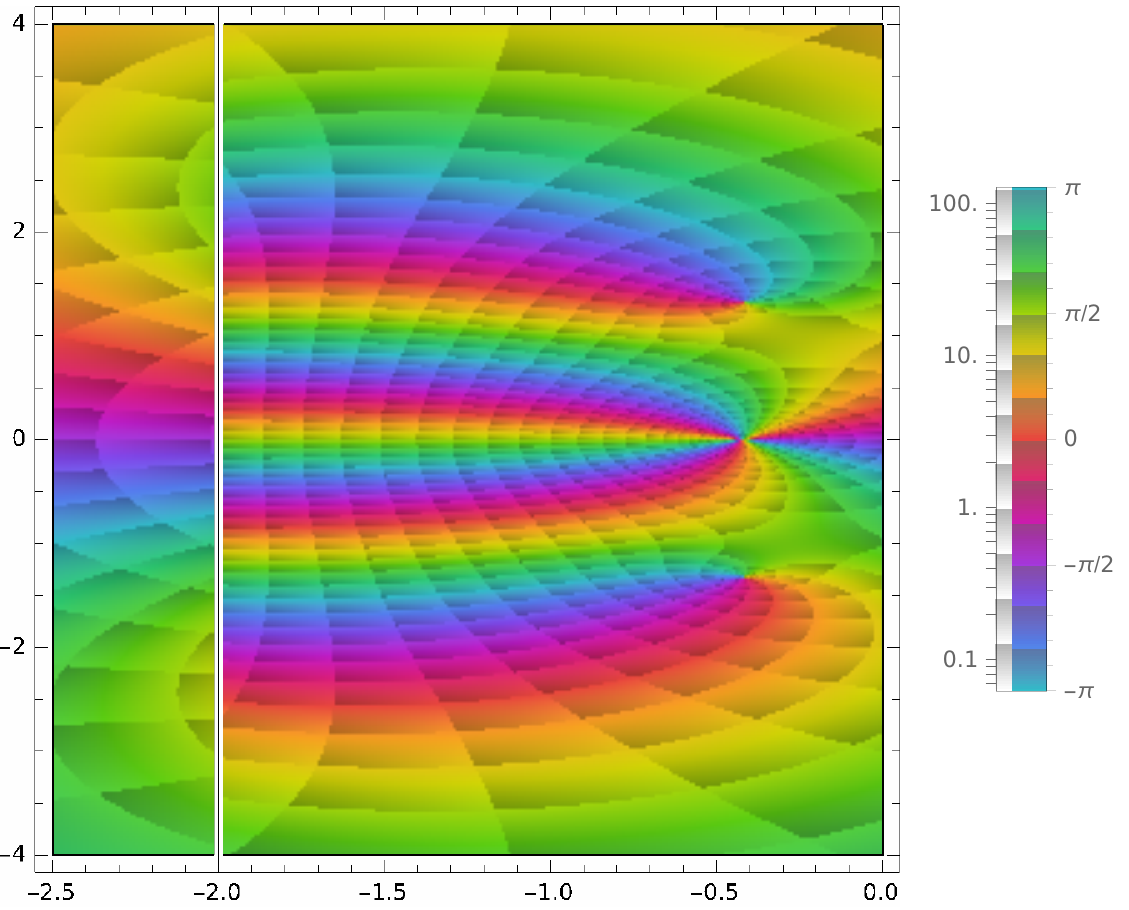}
		\caption{$ k=1$}
		\label{p1}
	\end{subfigure}%
	\begin{subfigure}{.5\textwidth}
		\centering
		\includegraphics[width=1\linewidth]{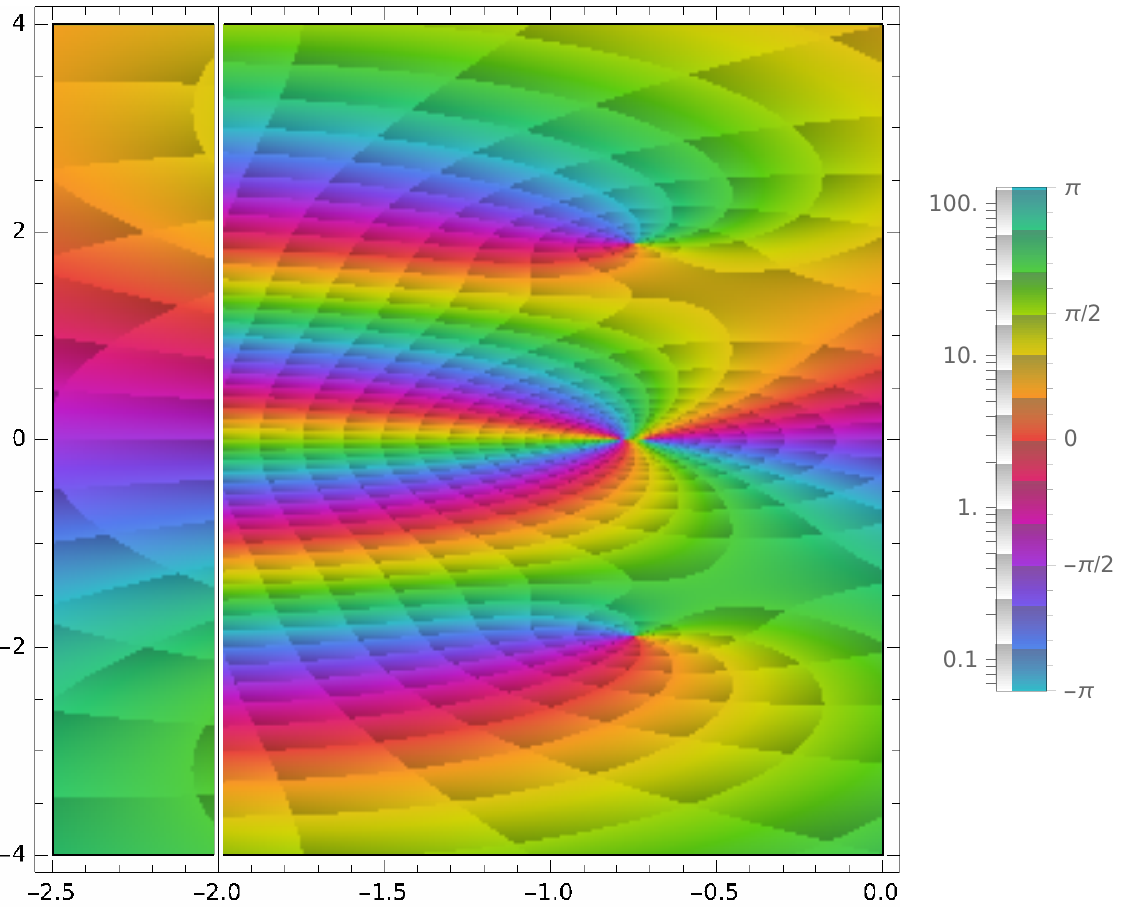}
		\caption{$ k=\sqrt{2}$}
		\label{ps2}
	\end{subfigure}
	\caption{Argument plot of the spectral function \eqref{detexpr} for $\tau=0.5$ and different values of $ k$. The zeros of the function \eqref{detexpr} in the complex plane define eigenvalues of the linearized BGK operator. These are points, where a small, counter-clockwise loop runs through the whole rainbow according to multiplicity.}
	\label{plotspec1}
\end{figure}

\begin{figure}
	\begin{subfigure}{.5\textwidth}
		\centering
		\includegraphics[width=1\linewidth]{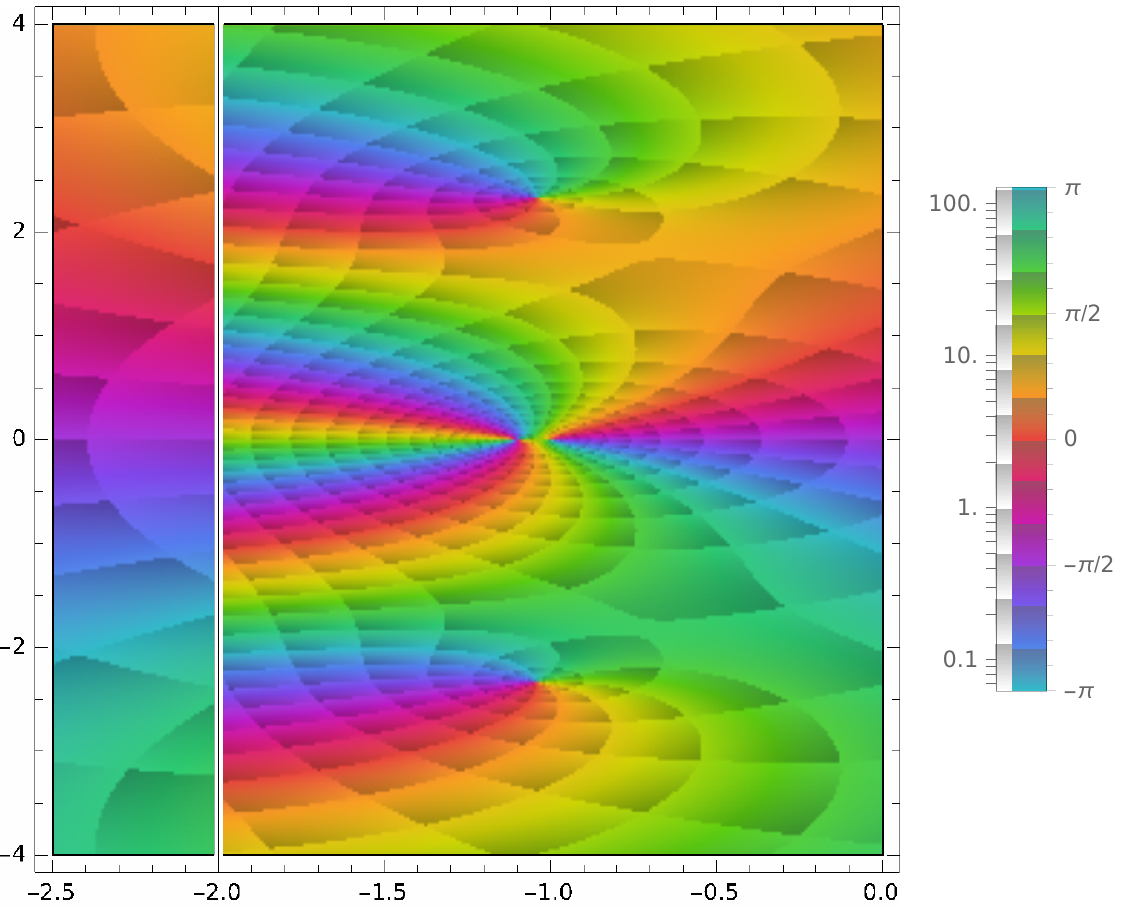}
		\caption{$ k=\sqrt{3}$}
		\label{ps3}
	\end{subfigure}%
	\begin{subfigure}{.5\textwidth}
		\centering
		\includegraphics[width=1\linewidth]{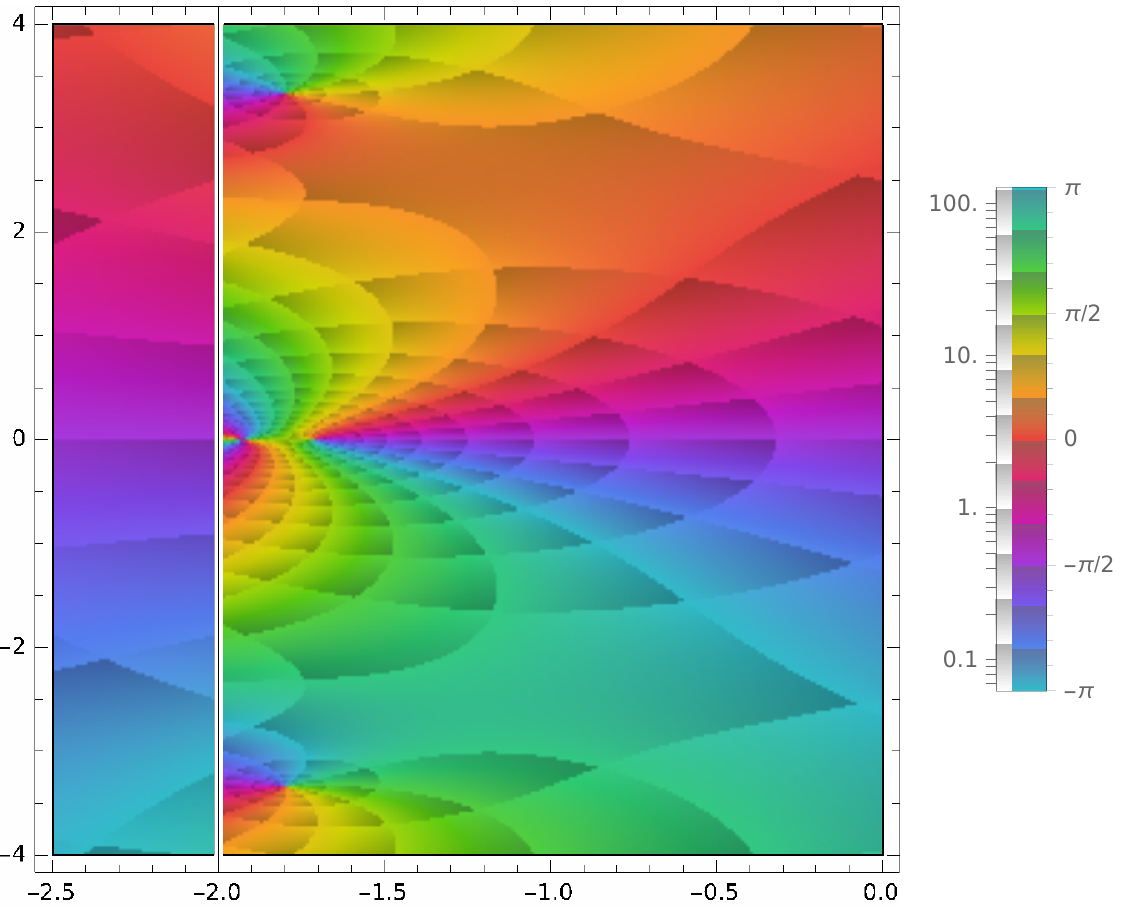}
		\caption{$ k=\sqrt{6}$}
		\label{ps6}
	\end{subfigure}
	\caption{Argument plot of the spectral function \eqref{detexpr} for $\tau=0.5$ and different values of $ k$. The zeros of the function \eqref{detexpr} in the complex plane define eigenvalues of the linearized BGK operator. These are points, where a small, counter-clockwise loop runs through the whole rainbow according to multiplicity. As we approach the critical wave number, the zeros move closer and closer to the essential spectrum ($\Re\lambda=-\frac{1}{\tau}$)}. 
		\label{plotspec2}
\end{figure}
\begin{figure}
	\begin{subfigure}{.5\textwidth}
		\centering
		\includegraphics[width=1\linewidth]{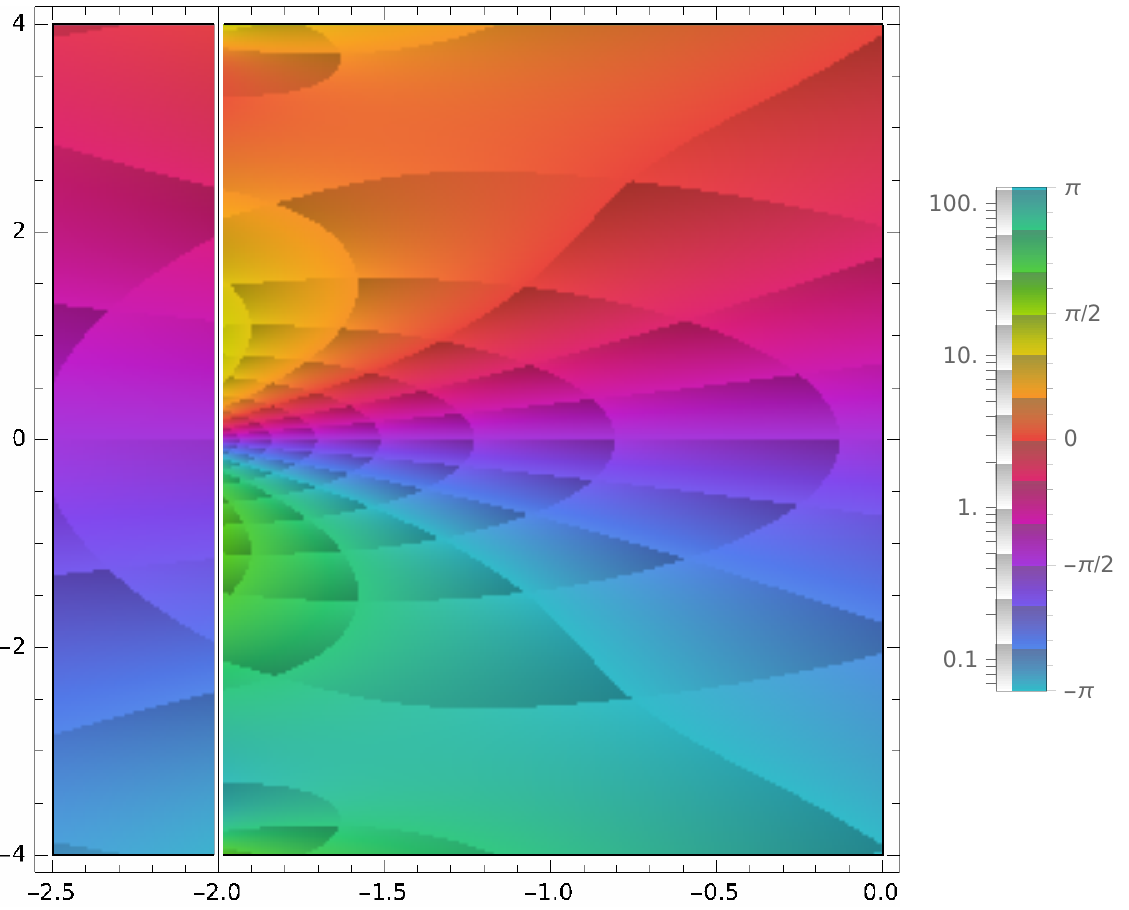}
		\caption{$ k=\sqrt{8}$}
		\label{ps8}
	\end{subfigure}%
	\begin{subfigure}{.5\textwidth}
		\centering
		\includegraphics[width=1\linewidth]{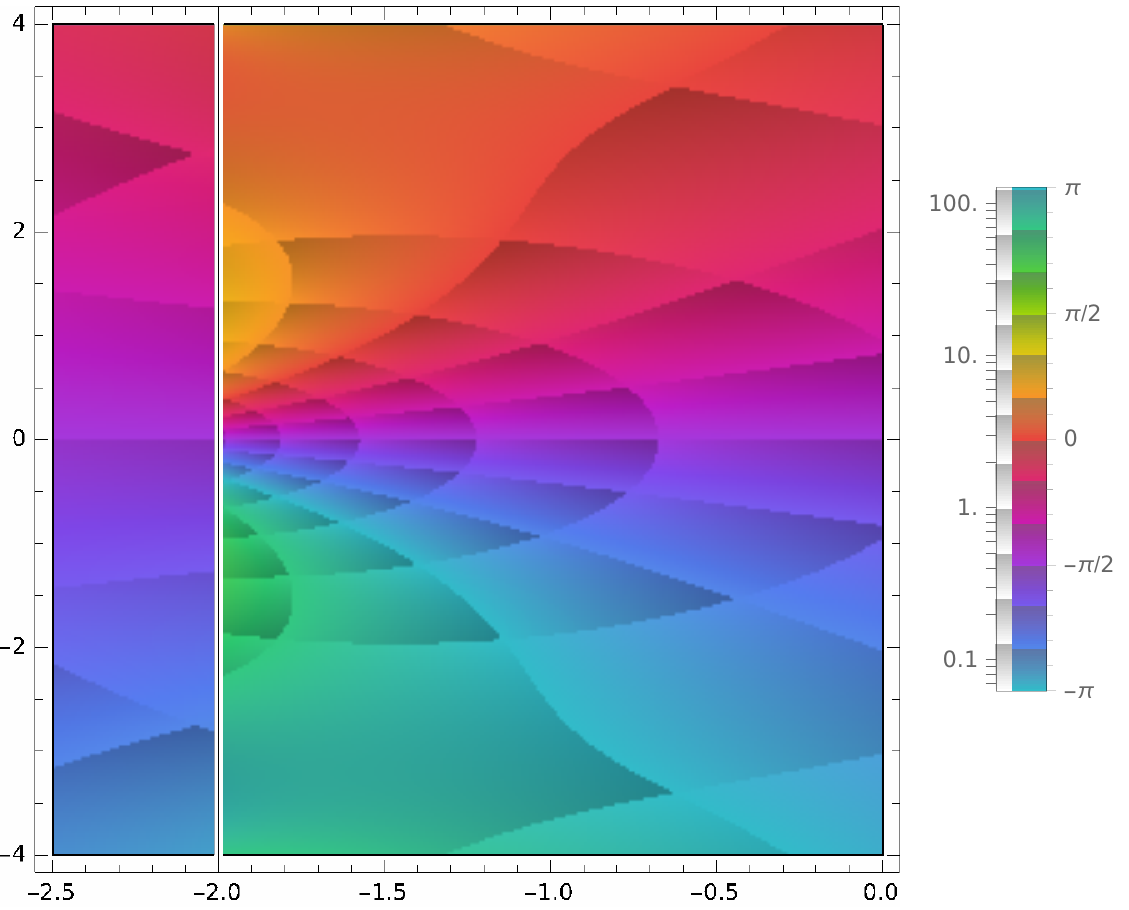}
		\caption{$ k=3$}
		\label{p3}
	\end{subfigure}
	\caption{Argument plot of the spectral function \eqref{detexpr} for $\tau=0.5$ and different values of $ k$. The zeros of the function \eqref{detexpr} in the complex plane define eigenvalues of the linearized BGK operator. These are points, where a small, counter-clockwise loop runs through the whole rainbow according to multiplicity. Since the wave number is above $k_{\rm crit}$, there exist, indeed, no zeros.}
		\label{plotspec3}
\end{figure}
\end{centering}

\begin{proof}
First, let us recall that any discrete eigenvalue $\lambda$ of $ \mathcal{L}_{\mathbf{k}}$ (and hence of $\mathcal{L}$) satisfies
\begin{equation}\label{boundlambda}
    -\frac{1}{\tau}<\Re\lambda\leq 0,
\end{equation}
by \eqref{dissipative}, which we will assume henceforth (of course, it would in fact follow from a slightly more detailed analysis of the following).
Since $\lambda$ and $\zeta$ are related by
\begin{equation}\label{lambdazeta}
 \lambda= -\frac{\ri  k \tau \zeta+1}{\tau},   
\end{equation}
this implies that $\Re\lambda =  k\Im\zeta-\frac{1}{\tau}$ and consequently
\begin{equation}
    0< \Im \zeta \leq \frac{1}{\tau k}.
\end{equation}
Our strategy is to apply Rouch{\'e}'s theorem to the function $\Gamma_{\tau k}$ by splitting it into a dominant part plus an (asymptotically) small part. To this end, we can focus on the family of rectangles $\mathbf{R}_a=\{- a, a, a+\ri\frac{1}{\tau k},-a+\ri\frac{1}{\tau k} \}$ for $a>0$. First, let us consider the asymptotics of $\Gamma_{\tau k}$ in $\zeta$ for fixed $\tau k$. \\

Since we are focused on the upper half-plane, we can consider $Z_{+}$ defined in \eqref{defIplus} as an analytic continuation together with its limit on the real line. In particular, we see from the asymptotics \eqref{Ipsymptotic} that
\begin{equation}\label{asyGamma}
    \begin{split}
    \Gamma_{\tau k}(\zeta) & = \frac{1}{6}\left(\zeta+6 \ri  k^3 \tau ^3-\zeta  (\zeta^2+5)  k^2 \tau ^2+2 \ri (\zeta ^2+3)  k \tau \right.\\
&\qquad\left.+Z(\zeta ) (\zeta ^2-(\zeta^4+4 \zeta ^2+11)  k^2 \tau ^2+2 \ri \zeta ^3  k \tau -5)-4 \ri Z^2 (\zeta )((\zeta ^2+1)  k \tau -\ri \zeta ) \right)\\
&\sim \frac{1}{6}\left(\zeta+6 \ri  k^3 \tau ^3-\zeta  (\zeta^2+5)  k^2 \tau ^2+2 \ri (\zeta ^2+3)  k \tau \right.\\
&\qquad-\sum_{n=0}^\infty \frac{(2n-1)!!}{\zeta^{2n+1}} (\zeta ^2-(\zeta^4+4 \zeta ^2+11)  k^2 \tau ^2+2 \ri \zeta ^3  k \tau -5)\\
&\qquad\left.-4 \ri \left( -\sum_{n=0}^\infty \frac{(2n-1)!!}{\zeta^{2n+1}}\right)^2((\zeta ^2+1)  k \tau -\ri \zeta ) \right),
\end{split}
\end{equation}
for $|\arg(\zeta)|\leq \frac{\pi}{2}-\delta$,  $\zeta\to\infty $, which, after rearranging and regrouping higher-order terms in $\zeta^{-1}$, gives
\begin{equation}\label{asyGamma2}
\begin{split}
\Gamma_{\tau k}(\zeta) &\sim \frac{1}{6}\left(\zeta+6 \ri  k^3 \tau ^3-\zeta  (\zeta^2+5)  k^2 \tau ^2+2 \ri (\zeta ^2+3)  k \tau \right.\\
&\qquad-(\zeta^{-1}+\zeta^{-3}) (\zeta ^2-(\zeta^4+4 \zeta ^2+11)  k^2 \tau ^2+2 \ri \zeta ^3  k \tau -5)+\mathcal{O}(|\zeta|^{-1})\\
&\qquad\left.-4 \ri \zeta^{-2}((\zeta ^2+1)  k \tau -\ri \zeta ) \right)+\mathcal{O}(|\zeta|^{-2})\\
&\sim \frac{1}{6}\left(\zeta+6 \ri  k^3 \tau ^3- k^2 \tau ^2\zeta^3  -5 k^2 \tau ^2\zeta+2 \ri  k \tau\zeta ^2+6\ri  k \tau  \right.\\
&\qquad-\zeta + k^2 \tau ^2 \zeta^3+4 k^2 \tau ^2\zeta+11 k^2 \tau ^2\zeta^{-1}-2\ri k\tau \zeta^2-5\zeta^{-1}\\
&\qquad-\zeta^{-1} + k^2 \tau ^2 \zeta+4 k^2 \tau ^2\zeta^{-1}+11 k^2 \tau^2\zeta^{-2}-2\ri k\tau -5\zeta^{-3}\\
&\left.\qquad -4\ri k\tau-4\ri k\tau\zeta^{-2}-4\zeta^{-1}+\mathcal{O}(|\zeta|^{-1})\right)\\
&\sim \ri ( k\tau)^3+\mathcal{O}(|\zeta|^{-1}),
\end{split}
\end{equation}
for  $|\arg(\zeta)|\leq \frac{\pi}{2}-\delta, \zeta\to\infty$, for any real number $0<\delta\leq \frac{\pi}{2}$.

\begin{remark}
It is a quite remarkable property of the spectral function $\Gamma_{\tau k}$ \eqref{defchar} that all the polynomial terms (up to order four) cancel exactly with the negative-power terms in the asymptotic expansion \eqref{Ipsymptotic} to give a constant asymptotic value in the limit. This is due to a subtle fine-tuning of the numerical coefficients of the polynomials. This property also guarantees the existence of a critical wave number (and hence implies that there are only finitely many discrete eigenvalues above the essential spectrum). At the outset, it is by no means clear that the spectrum should exhibit this cancellation property. {Indeed, numerical investigations actually leave this question unanswered} \cite{karlin2008exact}.
\end{remark}

Let us start with estimating $\Gamma_{\tau k}-\ri ( k\tau)^3$ on the real line. Because $x\mapsto |\Gamma_{\tau k}(x)-\ri ( k\tau)^3|$ is an even function for $x\in\mathbb{R}$, we can focus on $x>0$. Since $\Gamma_{\tau k}(x)\to \ri ( k\tau)^3$  as $x\to\infty$, we know that $x\mapsto |\Gamma_{\tau k}(x)-\ri ( k\tau)^3|$ is bounded on the real line. Since $\Gamma_{\tau k}(x)-\ri ( k\tau)^3$ only contains powers of $ k$ up to order two, we know that there exists a $k_1>0$ such that
\begin{equation}\label{absGam1}
|\Gamma_{\tau k}(x)-\ri ( k\tau)^3|<( k\tau)^3,
\end{equation}
for all $x\in\mathbb{R}$ and all $ k>k_1$. Indeed, expanding  $\Gamma_{\tau k}(x)-\ri(\tau k)^3 = \sum_{j=0}^2 \gamma_j(x)(\tau k)^j$ as a polynomial in $\tau k$ with $|\gamma_{j}(x)|\leq \tilde{\gamma}_j$ by the boundedness of $\Gamma_{\tau k}(x)-\ri(\tau k)^3$, any quadratic polynomial can be bounded eventually by a cubic polynomial in modulus,
\begin{equation}
    |\Gamma_{\tau k}(x)-\ri ( k\tau)^3| \leq \sum_{j=0}^2 \tilde{\gamma}_j (\tau k)^j < (\tau k)^3,
\end{equation}
for $k>k_1$ for some $k_1>0$.
By the same token, we conclude that $x\mapsto |\Gamma_{\tau k}(x+\frac{\ri}{ k\tau})-\ri ( k\tau)^3|$, is bounded for 
$x\in\mathbb{R}$ since \eqref{asyGamma2} holds in a cone containing the real axis. Therefore, since again $\Gamma_{\tau k}(x+\frac{\ri}{ k\tau})-\ri ( k\tau)^3$ is bounded for $x\in\mathbb{R}$, there exists a $k_2>0$ such that
\begin{equation}\label{est}
\left|\Gamma_{\tau k}\left(x+\frac{\ri}{ k\tau}\right)-\ri ( k\tau)^3\right|<( k\tau)^3,
\end{equation}
for all $x\in\mathbb{R}$ and all $ k>k_2$. Along the same lines as for \eqref{absGam1}, we expand the left-hand side of \eqref{est} in a quadratic polynomial in $k$ and use boundedness to conclude the estimate.
Clearly, an estimate of the form \eqref{est}
for all $x\in\mathbb{R}$, $0\leq y \leq \frac{1}{\tau k}$ and $ k>k_3$ for some $k_3>0$ holds true by compactness and the decay properties of $\Gamma_{\tau k}$. This shows that, for $ k$ large enough, we can bound the function $\Gamma_{\tau k}-\ri ( k\tau)^3$ on the rectangle $\mathbf{R}_a$ for any $a>0$ by the modulus of $\ri ( k\tau)^3$, which has no zeros in the strip at all (in particular, not in the strip $0\leq \Im\zeta\leq\frac{1}{\tau k}$). For $ k$ large enough, Rouch{\'e}'s theorem then implies that $\Gamma_{\tau k}$ cannot have any zeros for $0\leq \Im\zeta\leq\frac{1}{\tau k}$ either.\\
This proves the claim.
\end{proof}

\begin{centering}
\begin{figure}
	\begin{subfigure}{.5\textwidth}
		\centering
		\includegraphics[width=.8\linewidth]{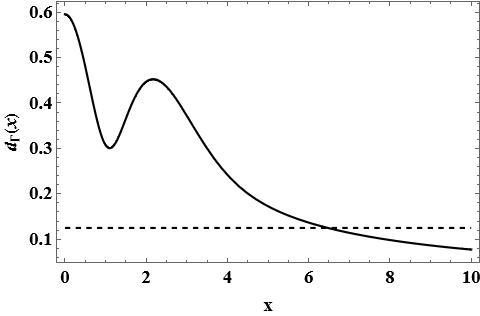}
		\caption{On the real line}
		\label{bound1}
	\end{subfigure}%
	\begin{subfigure}{.5\textwidth}
		\centering
		\includegraphics[width=.8\linewidth]{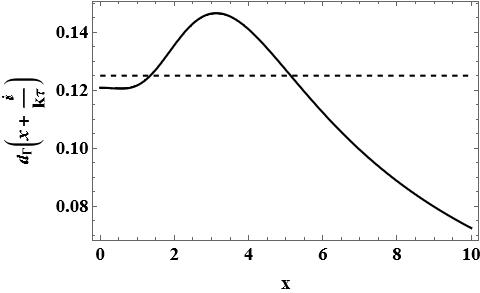}
		\caption{For $\Im \zeta =\frac{1}{\tau k}$}
		\label{bound1up}
	\end{subfigure}
	\caption{The function $\zeta\mapsto d(\zeta) = |\Gamma_{\tau k}(\zeta)-\ri( k\tau)^3|$ on the real line and on the line $\Im \zeta =\frac{1}{\tau k}$ for $\tau=0.5$ and $ k=1$ (solid lines) compared to $( k\tau)^3$ (dashed lines).}
	\label{bounds1}
\end{figure}

\begin{figure}
	\begin{subfigure}{.5\textwidth}
		\centering
		\includegraphics[width=.8\linewidth]{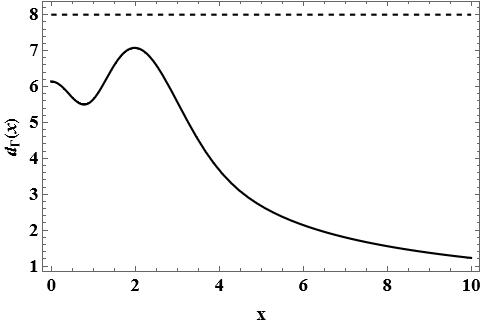}
		\caption{On the real line}
		\label{bound2}
	\end{subfigure}%
	\begin{subfigure}{.5\textwidth}
		\centering
		\includegraphics[width=.8\linewidth]{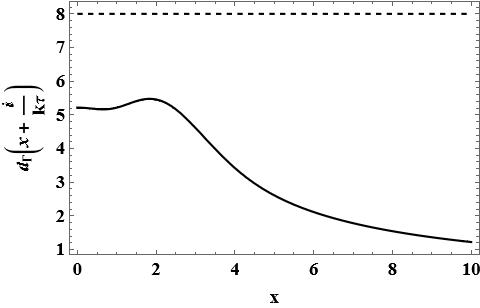}
		\caption{For $\Im \zeta =\frac{1}{\tau k}$}
		\label{bound2up}
	\end{subfigure}
	\caption{The function $\zeta\mapsto d(\zeta) = |\Gamma_{\tau k}(\zeta)-\ri( k\tau)^3|$ on the real line and on the line $\Im \zeta =\frac{1}{\tau k}$ for $\tau=0.5$ and $ k=4$ (solid lines) compared to $( k\tau)^3$ (dashed lines).}
	\label{bounds2}
\end{figure}
\end{centering}
Now, let us prove that
\begin{equation}
\tilde{\Gamma}_{\tau k}(\lambda):=\frac{1}{(\ri k\tau)^{3}}\Gamma_{\tau k}\left(\frac{-\tau\lambda-1}{\ri k\tau}\right)
\end{equation}
has exactly three zeros (one real, two complex conjugate, which we will prove later) for $ k$ small enough.
\begin{proof}
To this end, we again use the asymptotic expansion \eqref{Ipsymptotic} up to order three for the limit $ k\to 0$ together with expansion similar to those derived in \eqref{asyGamma} and \eqref{asyGamma2}:
\begin{equation}
\begin{split}
\textcolor{blue}{\tilde{\Gamma}_{\tau k}}(\lambda) & \sim \frac{1}{6(\ri k\tau)^{3}}\left(\zeta+6 \ri  k^3 \tau ^3-\zeta  (\zeta^2+5)  k^2 \tau ^2+2 \ri (\zeta ^2+3)  k \tau \right.\\
&\qquad+Z(\zeta ) (\zeta ^2-(\zeta^4+4 \zeta ^2+11)  k^2 \tau ^2+2 \ri \zeta ^3  k \tau -5)\\
&\qquad\left.-4 \ri Z^2 (\zeta )((\zeta ^2+1)  k \tau -\ri \zeta ) \right)\Big|_{\zeta=\frac{-\tau\lambda-1}{\ri k\tau}}\\
&\sim\frac{1}{6(\ri k\tau)^{3}}\left(\zeta+6 \ri  k^3 \tau ^3-\zeta  (\zeta^2+5)  k^2 \tau ^2+2 \ri (\zeta ^2+3)  k \tau \right.\\
&\qquad+(-\zeta^{-1}-\zeta^{-3}-3\zeta^{-5}+\mathcal{O}(|\zeta|^{-7})) (\zeta ^2-(\zeta^4+4 \zeta ^2+11)  k^2 \tau ^2+2 \ri \zeta ^3  k \tau -5)\\
&\qquad\left.-4 \ri (-\zeta^{-1}-\zeta^{-3}-3\zeta^{-5}+\mathcal{O}(|\zeta|^{-7}))^2 ((\zeta ^2+1)  k \tau -\ri \zeta ) \right)\Big|_{\zeta=\frac{-\tau\lambda-1}{\ri k\tau}},
\end{split}
\end{equation}
which, after plugging in the transformation \eqref{lambdazeta}, gives
\begin{equation}
\begin{split}
\tilde{\Gamma}_{\tau k}(\lambda) &\sim\frac{1}{6(\ri k\tau)^{3}}\left[\mathcal{O}(|\zeta|^{-3})( k\tau)^2+6\ri( k\tau)^3+( k\tau)^2\left(18 \zeta^{-1}+23 \zeta^{-3}+33\zeta^{-5}\right)\right.\\
&\qquad\left.-2\ri  k \tau \left(9 \zeta^{-2}+18 \zeta^{-4}+26 \zeta^{-6}+30 \zeta^{-8}+18\zeta^{-10}\right)-\left(6 \zeta^{-3}+13 \zeta^{-5}+24 \zeta^{-7}+36\right)\right]\Big|_{\zeta=\frac{-\tau\lambda-1}{\ri k\tau}}\\
&\sim \frac{1}{6(\ri k\tau)^{3}}\left[6\ri( k\tau)^3+18\ri( k\tau)^3(-\tau\lambda-1)^{-1}-18(\ri k\tau)^3(-\tau\lambda-1)^{-2}\right.\\
&\qquad\left.-6(\ri k\tau)^3(-\tau\lambda-1)^{-3}+\mathcal{O}( k^4)\right]\\
&\sim -\frac{\lambda^3}{(\lambda \tau+1)^3}+\mathcal{O}( k),
\end{split}
\end{equation}
i.e., in the limit $ k\to 0$, the spectral function \eqref{defchar} has a triple zero at $\lambda=0$. The cubic scaling in $ k$ in front of the above expression cancels exactly with the terms inside the bracket, leaving only the term $\lambda^3$ in the limit $ k\to 0$.  This is consistent with the spectrum of $ \mathcal{L}_0$ containing zero as an isolated eigenvalue, see \eqref{sigma0}. By continuity of the spectrum, this implies that the there will emanate exactly three discrete eigenvalues as zeros of the spectral function $\Gamma_{\tau k}$. 
\end{proof}

\subsection{Hydrodynamic Modes and their Corresponding Critical Wave Numbers}\label{spectral4}

\begin{figure}
		\centering
		\includegraphics[width=.5\linewidth]{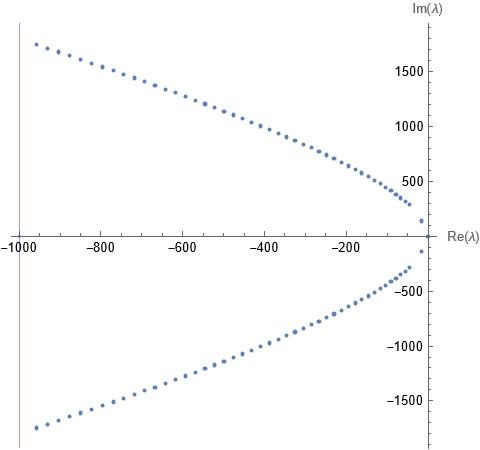}
		\caption{The acoustic modes for $\tau=0.001$ and wave numbers up to the critical wave number together withe the vertical line $\Re\lambda=-\frac{1}{\tau}$}
		\label{Acoustic}
\end{figure}

Now, let us take a closer look at the eigenvalues. From \eqref{detexpr}, it follows immediately that there exists a sequence of real eigenvalue of algebraic multiplicity two which we call \textit{shear mode} and denote as $ k\mapsto \lambda_{\rm shear}( k)$.\\
A closer look at \eqref{defchar} reveals that the function $\Gamma_{\tau k}$ maps imaginary numbers to imaginary numbers (since also $Z|_{\ri\mathbb{R}}\subseteq \ri\mathbb{R}$ by \eqref{g0}). As a consequence, $\tilde{\Gamma}_{\tau k}(\lambda)$ maps real numbers to real numbers. This shows that, together with the above considerations, that, for each wave number small enough, there exists exactly one real zero and two complex conjugated zeros. As a consequence of the symmetry property in Remark \ref{remsym}, we even know that any eigenvalue is either real or comes in a complex conjugated pair.\\
Consequently, apart from the shear mode, there exists a sequence of pairs of complex conjugated eigenvalues which we call \textit{acoustic modes} and denote as $ k\mapsto \lambda_{\rm ac}( k)$ and $ k\mapsto \lambda_{\rm ac}^*( k)$. Figure \ref{Acoustic} shows the distribution of acoustic modes for a given relaxation time and varying wave number.\\
Furthermore, there exists another simple, real eigenvalue called \textit{diffusion mode} which we denote as $ k\mapsto \lambda_{\rm diff}( k)$. Each mode has its own critical wave number. In conclusion, the spectrum is given by
\begin{equation}\label{sigmaLk}
\sigma({ \mathcal{L}_{\mathbf{k}}} )=\left\{-\frac{1}{\tau} +\ri\mathbb{R}\right\} \cup \{\lambda_{\rm shear}( k),\lambda_{\rm diff}( k),\lambda_{\rm ac}( k),\lambda_{\rm ac}^*( k)\},
\end{equation}
for $ k$ smaller than the respective critical wave number. \\

Before we turn to more specific estimates on the critical wave number of each eigenvalue branch, which will also indicate that, indeed, no additional branches can bifurcate out of the essential spectrum, let us prove that the four branches in \eqref{sigmaLk} cannot merge. To this end, we will need the following lemma: 
\begin{lemma}\label{nonneg}
    Let $k>0$. The $\zeta$-derivative of the spectral function \eqref{defchar} restricted to the positive imaginary axis is strictly positive.
\end{lemma}
Proof is given in Appendix \ref{app:lemma}.

As a consequence of Lemma \ref{nonneg}, it follows that the spectral function $\Gamma_{\tau k}$ cannot have degenerate roots on the real axis. Because of the symmetry property \eqref{symGam}, the pair of acoustic modes could only merge at the real axis through a twice-degenerate real root, which is excluded. Therefore, the acoustic modes have to have non-zero imaginary part for all $k$ where they exist. By the symmetry property, the real diffusion mode stays real for all wave number where it exists. 
\begin{remark}
    The proof above rules out the possibility that the pair of acoustic modes merges (at the real axis) to produce a twice-degenerated real eigenvalue. For other kinetic models, however, such as e.g. the Shakhov model, this possibility \textit{cannot} in fact  be excluded. Indeed, as proven in \cite{kogelbauer2024spectral}, for a certain value of Prandtl number, two diffusion modes can collide at the real axis to produce another pair of acoustic modes called \textit{second sound}. 
\end{remark}

\begin{remark}
We note that the eigenvalues (and hence the spectrum) depends on wave number only through $\tau k$. This implies that, while the eigenvectors depend on the full wave vector $\mathbf{k}$, the form of the spectrum only depends on the dimensionless parameter $\tau k$ and the existence of the hydrodynamic manifold (as a linear combination of eigenvectors) is independent of the relaxation time. If the relaxation time decreases, the critical wave number of each mode is increased, thus allowing for more eigenvalues in each family of modes. Consequently, decreasing the relaxation time increases the (finite) \textit{dimension} of the hydrodynamic manifold.\\

In the limit $\tau\to 0$, the eigenvalues accumulate at the essential spectrum and we cannot separate a hydrodynamic manifold any longer, since the corresponding spectral projection does not exist (no closed contour can be defined that encircles the set of discrete eigenvalues, while not intersecting the essential spectrum). 
\end{remark}

\begin{figure}
		\centering
		\includegraphics[width=.7\linewidth]{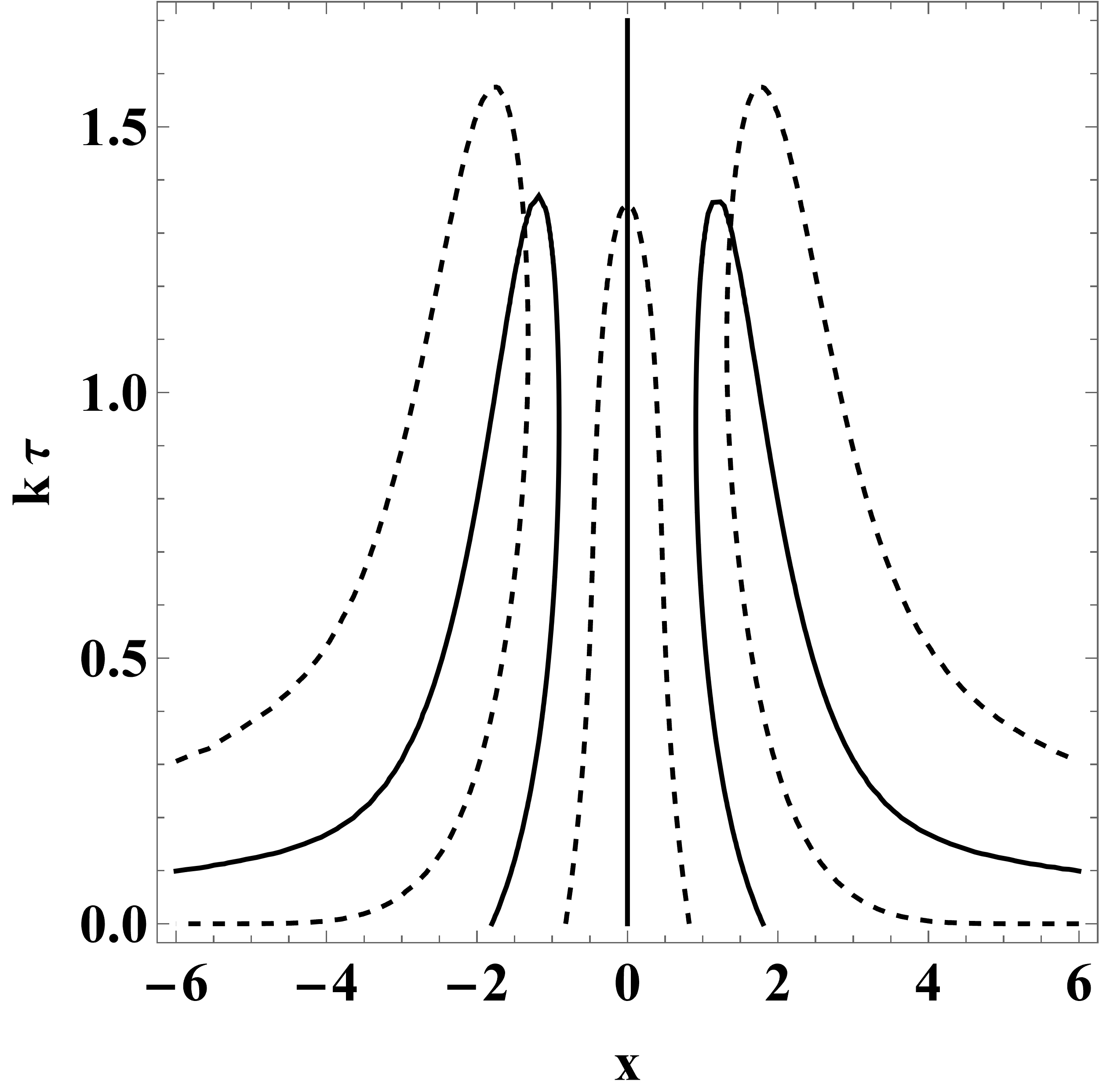}
		\caption{The zero sets of equations \eqref{critacoustic}. The intersection of the solid line ($\Re \Gamma_{\tau k}(x)$) with the dashed line ($\Im \Gamma_{\tau k}(x)$) gives the critical wave numbers for the acoustic modes (and the diffusion mode on the real line as well).}
		\label{ZeroSet}
\end{figure}

To finish the spectral analysis, let us derive some information about the critical wave number of the four hydrodynamic modes. Since $|Z|\leq \sqrt{\frac{\pi}{2}}$ with equality exactly at zero (continuously extended from both sides), we immediately conclude that
\begin{equation}
k_{\rm crit}(\lambda_{\rm shear})= \sqrt{\frac{\pi}{2}}\frac{1}{\tau}\approx 1.25331\frac{1}{\tau}. 
\end{equation}
from equation \eqref{spec2}. This is consistent with the result obtained in \cite{kogelbauer2021} (equation (2.53) in \cite{kogelbauer2021}).\\
Since the diffusion mode is real, and wanders from zero to $-\frac{1}{\tau}$ as $ k$ increases, we can recover the critical wave number by taking the limit $\lambda\to-\frac{1}{\tau}$ (on the branch $Z_+$) in \eqref{defchar}. Since $\lim_{\zeta\to 0, \Im\zeta>0} Z(\zeta)=\ri\sqrt{\frac{\pi}{2}}$, we obtain the critical wave number $k_{\rm crit}(\lambda_{\rm diff})$ as a zero of the cubic polynomial
\begin{equation}
6(k\tau)^3-11\sqrt{\frac{\pi}{2}}(k\tau)^2+\left(6+2\pi\right)k\tau-5\sqrt{\frac{\pi}{2}}=0.
\end{equation}
The only real solution is approximately given by
\begin{equation}
k_{\rm crit}(\lambda_{\rm diff})\approx 1.35603\frac{1}{\tau}.
\end{equation}
Now, let us turn to the acoustic mode. We know that at the critical wave number, the two complex conjugated acoustic modes will merge into the essential spectrum. This happens when $\Re\lambda=-\frac{1}{\tau}$. So, let us assume that $\lambda=-\frac{1}{\tau}-\ri k x$, which amount to setting $\zeta=x$ in \eqref{defchar}. We obtain two equations (real and imaginary part of $\Gamma_{\tau k}(x)$):
\begin{equation}\label{critacoustic}
\begin{split}
&\frac{1}{12} e^{-x^2} \left(\erfi\left(\frac{x}{\sqrt{2}}\right) \left(\sqrt{2 \pi } (\tau k)^2 e^{\frac{x^2}{2}} \left(x^4+4 x^2+11\right)-8 \pi  \tau k \left(x^2+1\right)-\sqrt{2 \pi } e^{\frac{x^2}{2}} \left(x^2-5\right)\right)\right.\\
&\qquad\left.-4 \pi x \erfi\left(\frac{x}{\sqrt{2}}\right)^2-2 x\left(e^{x^2} \left((\tau k)^2\left(x^2+5\right)-1\right)+\sqrt{2 \pi } \tau ke^{\frac{x^2}{2}} x^2-2 \pi \right)\right)=0,\\
&\frac{1}{12} e^{-x^2} \left(-4 \pi  \tau k \left(x^2+1\right)
   \erfi\left(\frac{x}{\sqrt{2}}\right)^2+\erfi\left(\frac{x}{\sqrt{2}}\right) \left(8 \pi  x-2 \sqrt{2 \pi
   } \tau k e^{\frac{x^2}{2}} x^3\right)\right.\\
   &\qquad+4 \tau k e^{x^2} \left(3
   \tau k^2+x^2+3\right)\\
   &\qquad\left.+\sqrt{2 \pi } e^{\frac{x^2}{2}}
   \left(-\left((\tau k)^2 \left(x^4+4
   x^2+11\right)\right)+x^2-5\right)+4 \pi  \tau k
   \left(x^2+1\right)\right)=0,
\end{split}
\end{equation}
for $x\in\mathbb{R}$. The zero sets of equations \eqref{critacoustic} are shown in Figure \ref{ZeroSet}.

Solving system \eqref{critacoustic} numerically gives the following approximation for the critical wave number of the acoustic mode:
\begin{equation}
k_{\rm crit}(\lambda_{\rm ac})=k_{\rm crit}(\lambda_{\rm ac}^*)\approx 1.31176\frac{1}{\tau}. 
\end{equation}

\begin{remark}
The critical wave numbers obtained before depend inversely on the (non-dimensional) relaxation parameter. Transforming back to physical units, we see that the critical wave number is numerically proportional to the inverse mean-free path \eqref{defmeanfree}. Indeed, we obtain that
\begin{equation}
k_{\rm crit}\sim \sqrt{\frac{k_BT_0}{m}}\frac{1}{\tau_{\rm phys}}\sim \frac{1}{l_{\rm mfp}}. 
\end{equation}
\end{remark}

\begin{remark}
The structure of the zero set of \eqref{critacoustic} over $(x,k)$ - admitting only three zeros - also shows that there cannot bifurcate any additional eigenvalue branches from the essential spectrum (apart from the four hydrodynamic branches). Indeed, the acoustic and diffusive branches being analytic (and in particular continuous curves) that cannot cross, they have to intersect the boundary at exactly three points. 
  \end{remark}
  \begin{remark}
       Since the eigenvalue branches depend analytically on wave number (see also general discussion in \cite{ellis1975first}), the instantaneous direction in which a zero to $\Gamma_{k \tau}$ moves with increase of wave number is given by\begin{equation}    \frac{\partial \zeta}{\partial k} =-\frac{\partial_k \Gamma_{\tau k}(\zeta)}{\partial_{\zeta}\Gamma_{\tau k} (\zeta)},\end{equation}which follows from differentiating the relation $\Gamma_{\tau k}(\zeta(k)) = 0$, provided that $\partial_{\zeta}\Gamma_{\tau k} (\zeta)\neq 0$. Since the eigenvalues only move to the left (towards the essential spectrum), we have that
       \begin{equation}\label{ineq}
           -\Im \frac{\partial_k \Gamma_{\tau k}(\zeta(k))}{\partial_{\zeta}\Gamma_{\tau k} (\zeta(k))} <0,
       \end{equation}
       which gives an alternative justification why no additional eigenvalue branches can bifurcate out of the essential spectrum. We leave  the rigorous proof of inequality \eqref{ineq} as a conjecture. 
  \end{remark}

\section{Conclusion and Further Perspectives}

We have given a complete and (up to the solution of a transcendental equation) explicit description of the spectrum of the three-dimensional BGK equation linearized around a global Maxwellian. Further, we identified (and therefore confirmed) the existence of three families of modes (shear, diffusion and acoustic) and we gave an explicit description of critical wave numbers. The analysis allowed us to infer that the discrete spectrum consists of a finite number of eigenvalues, thus implying that the dispersion relation remains bounded also for the acoustic modes. In particular, we obtained explicit values for the critical wave numbers. In the second part \cite{kogelbauerBGKspectral2}, we will use the explicit knowledge of the hydrodynamic branches to construct a closure relation for the macroscopic variables. \\

Furthermore, the explicit knowledge of the spectral function \eqref{defchar} allows us to infer more refined approximations to the exact non-local hydrodynamics. This will involve expansions not in terms of relaxation time or wave number, but much rather in terms of the variable $\zeta$ in \eqref{defchar}. This could also improve present numerical methods
\cite{karlin2008exact}.\\

Finally, the spectral properties of the linear three-dimensional BGK equation will also serve as the basis for nonlinear analysis in terms of invariant manifolds. Indeed, the fact that the discrete spectrum is well separated from the essential spectrum allows us to define a spectral projection for the \textit{whole} set of eigenvalues, thus giving the first-order approximation (in terms of nonlinear deformations) to the hydrodynamic manifolds. In particular, we expect that the theory of thermodynamic projectors \cite{GORBAN2004391} may be helpful in proving the nonlinear extension.

\section*{Acknowledgement}
This work was supported by European Research Council (ERC) Advanced Grant no. 834763-PonD (F.K. and I.K.).
The authors would like to thank the anonymous reviewers for their useful comments and suggestions, especially in connection with the Weinstein--Aronszajn determinant.




\appendix

\section{Proof of Lemma \ref{nonneg}}\label{app:lemma}
\begin{figure}
    \centering
    \includegraphics[width=0.45\linewidth]{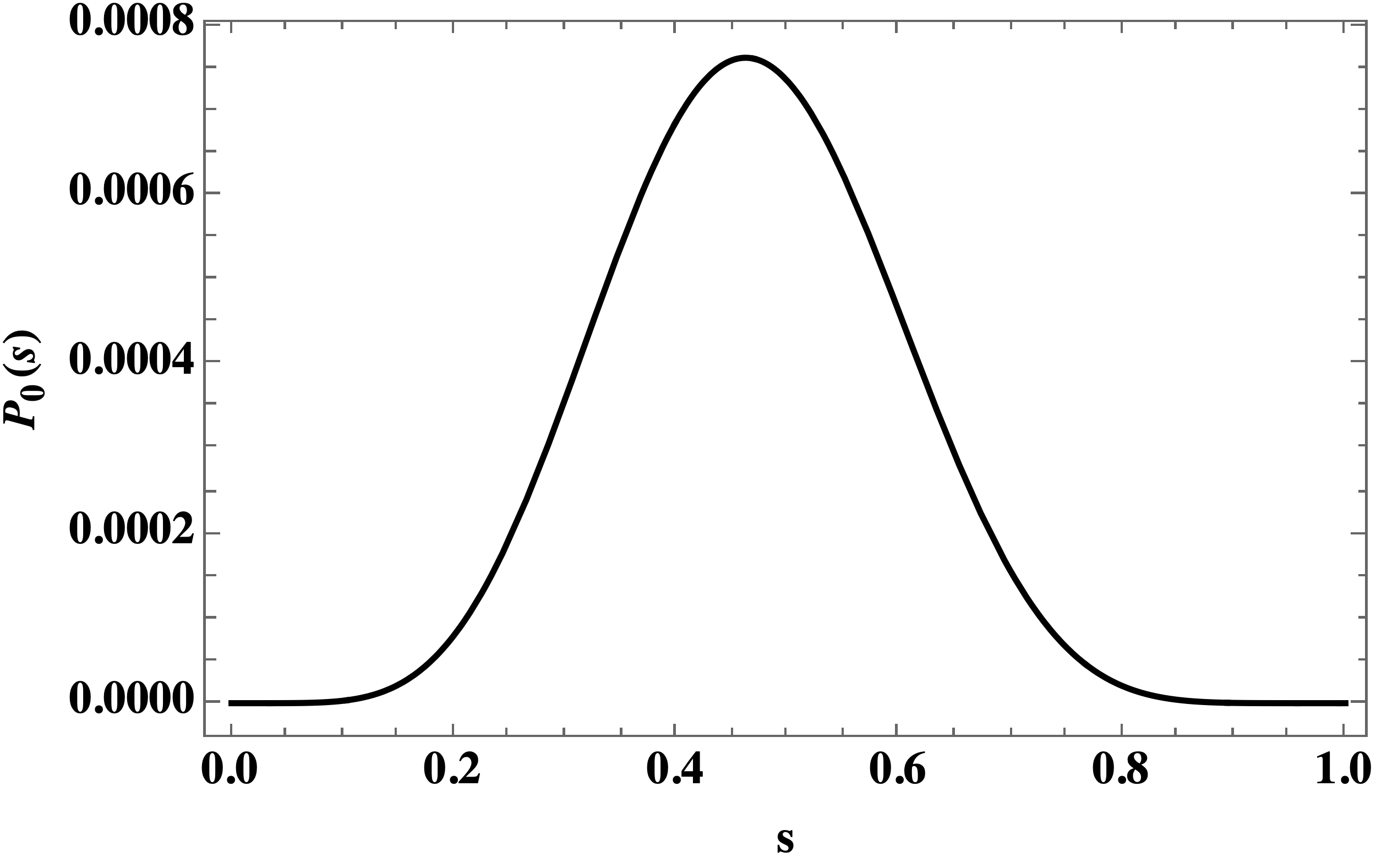}
    \includegraphics[width=0.45\linewidth]{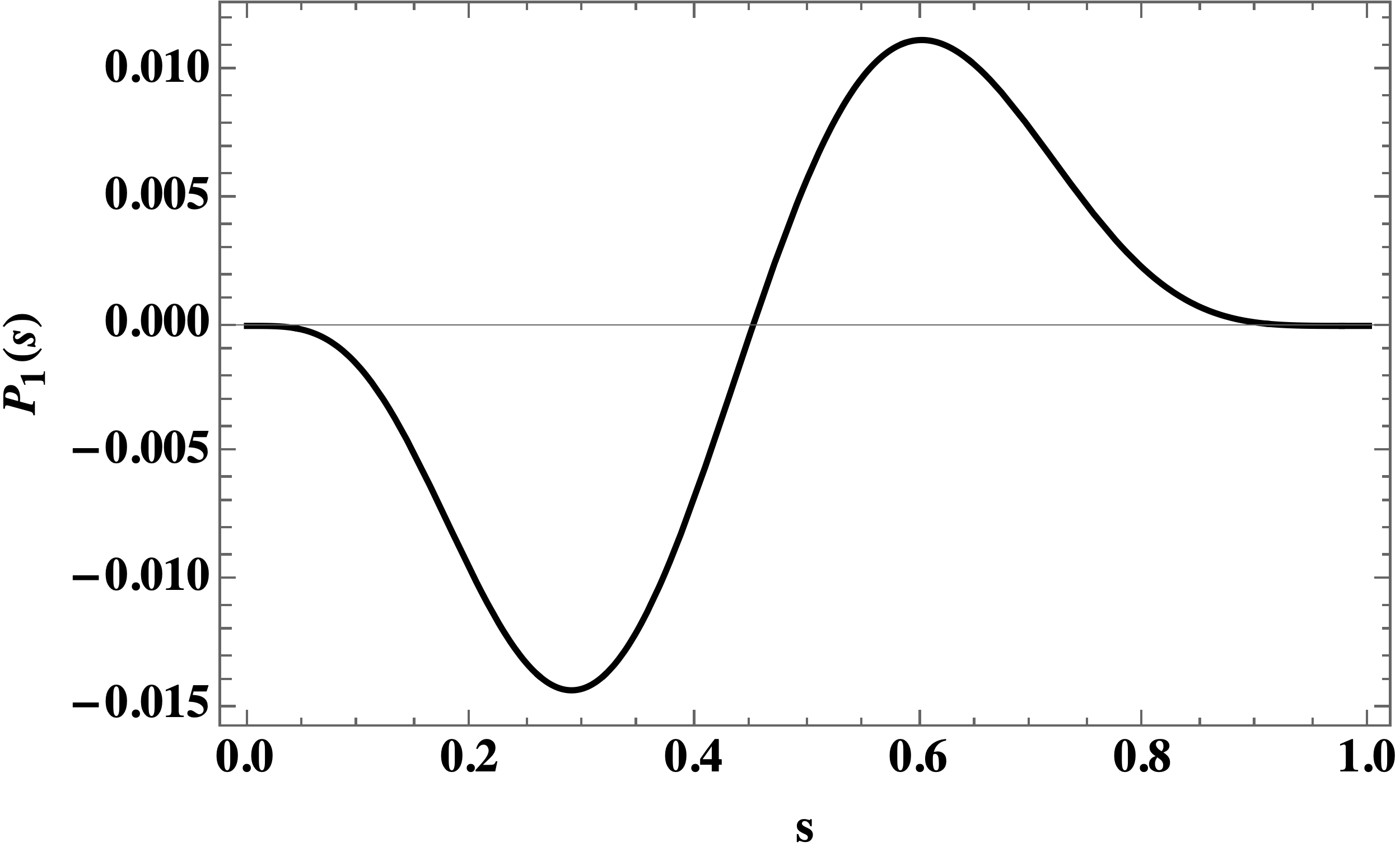}
    \includegraphics[width=0.45\linewidth]{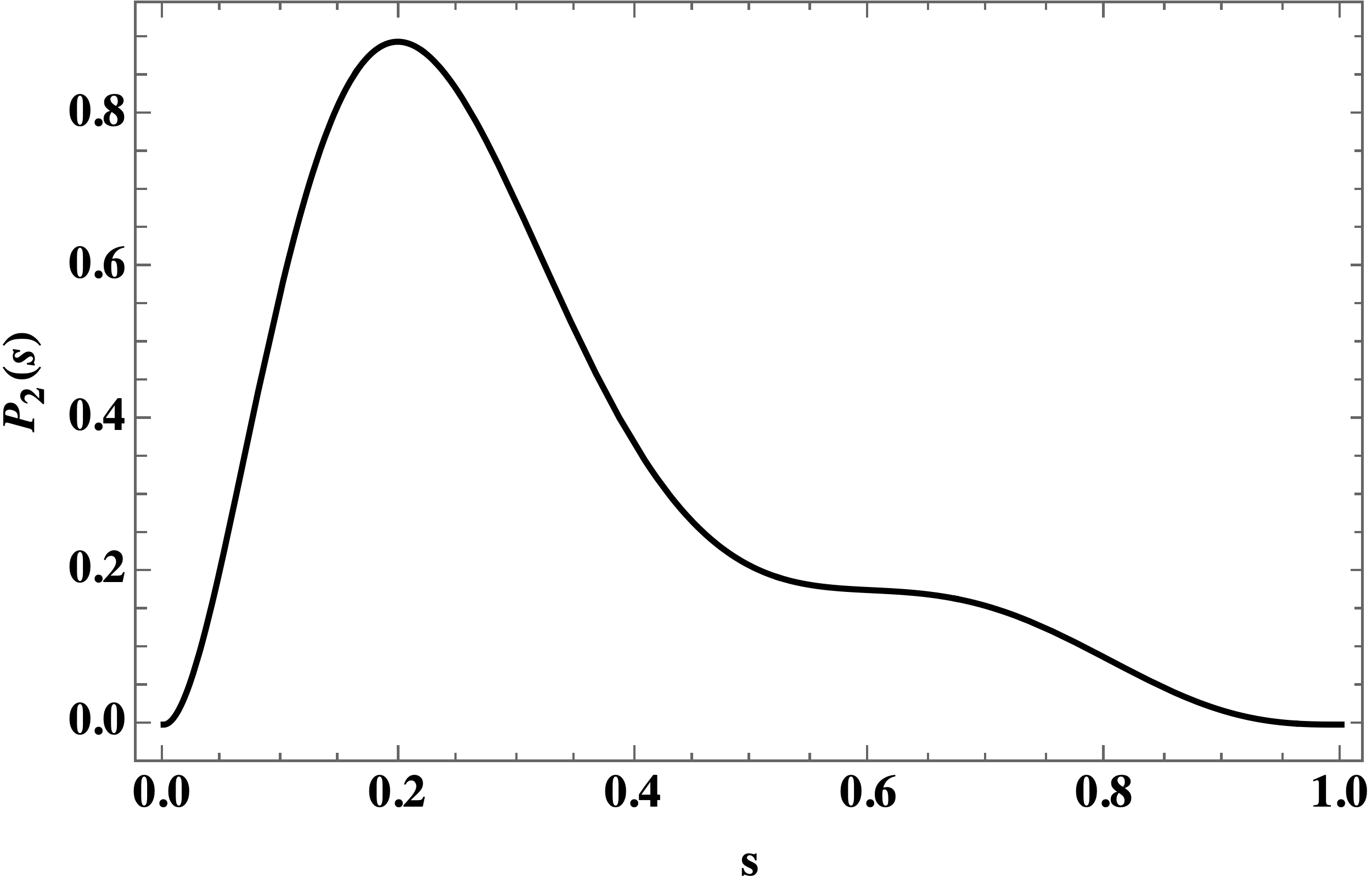}
    \includegraphics[width=0.45\linewidth]{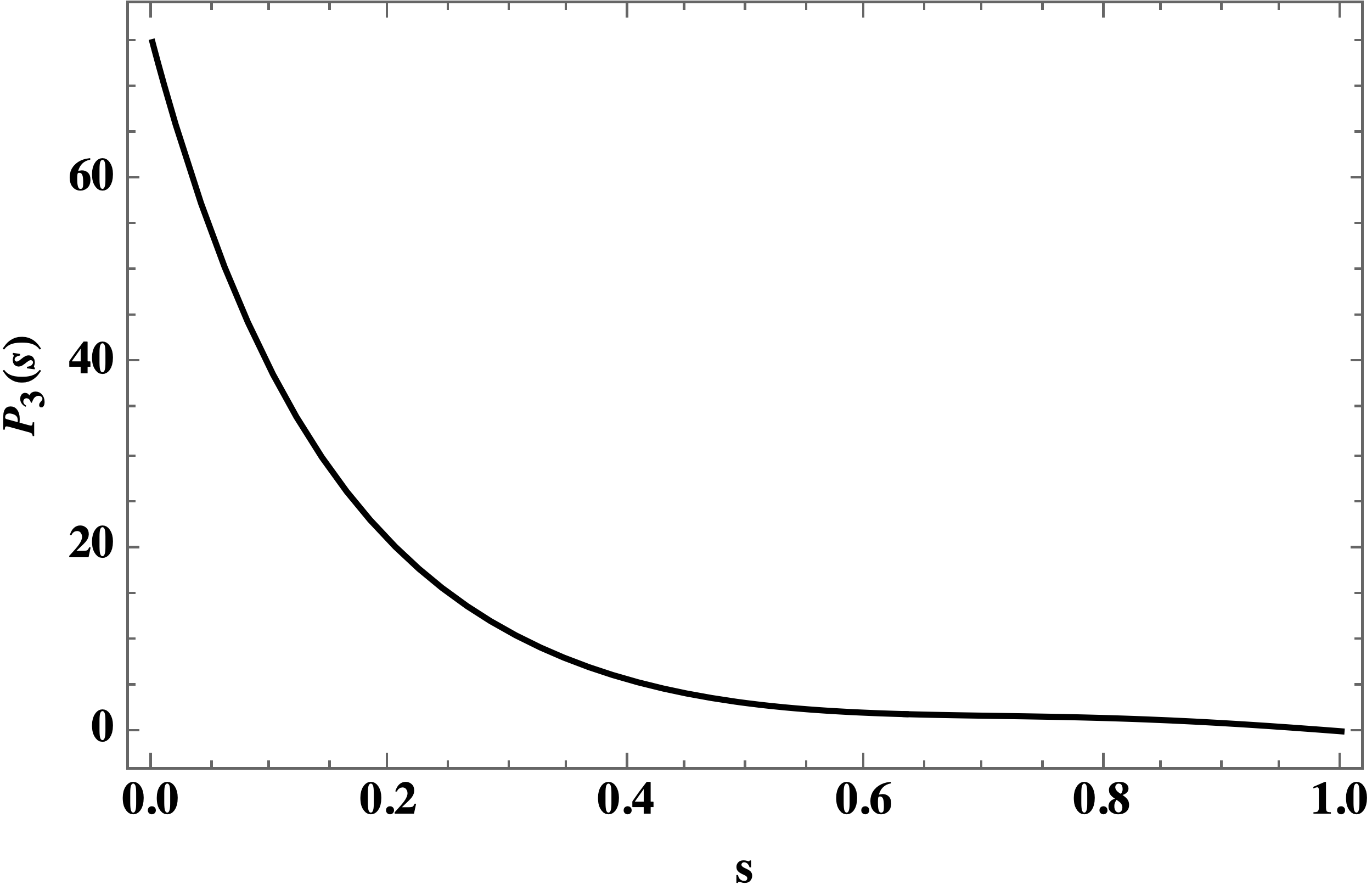}
    \includegraphics[width=0.45\linewidth]{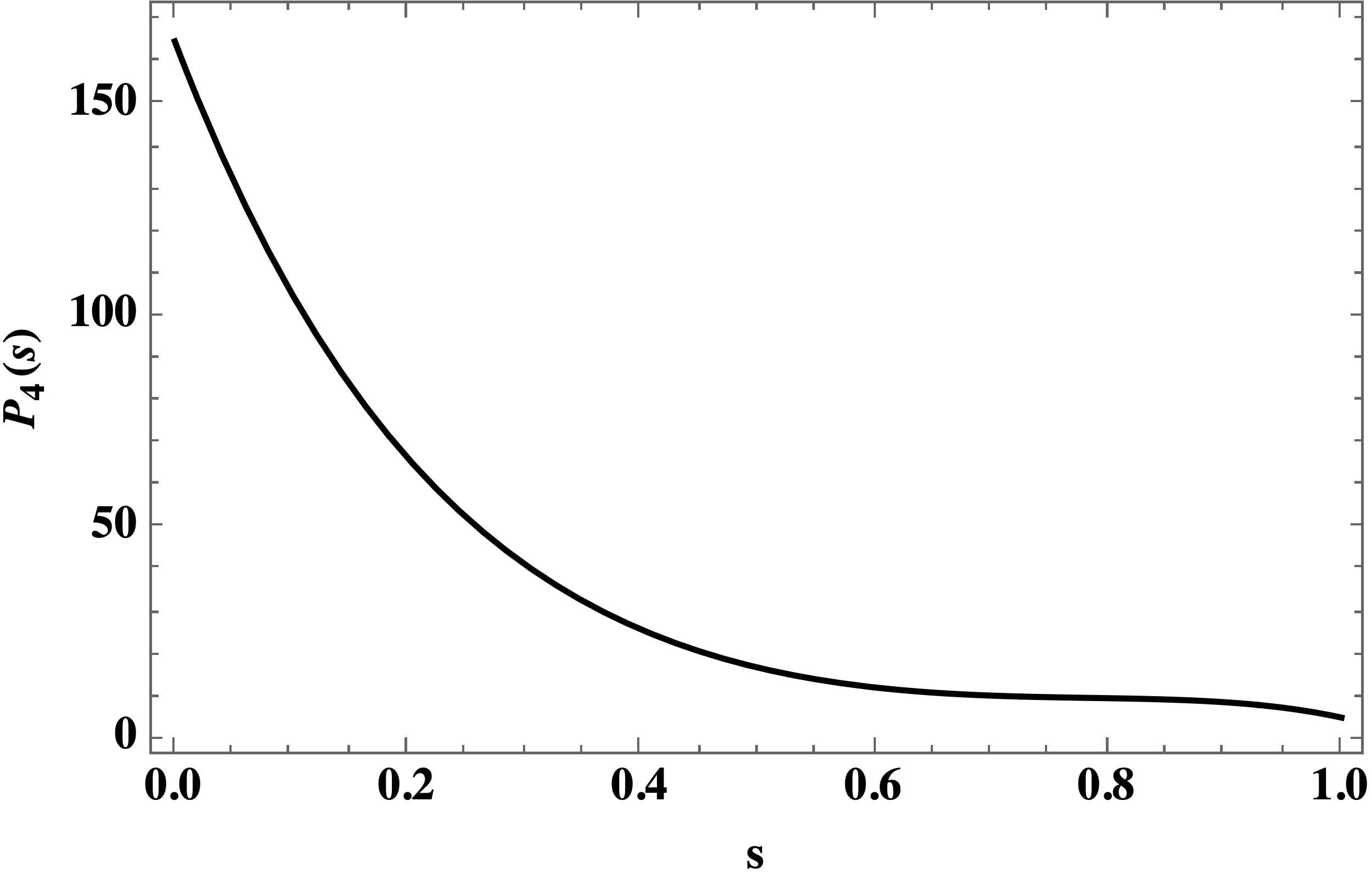}
    \includegraphics[width=0.45\linewidth]{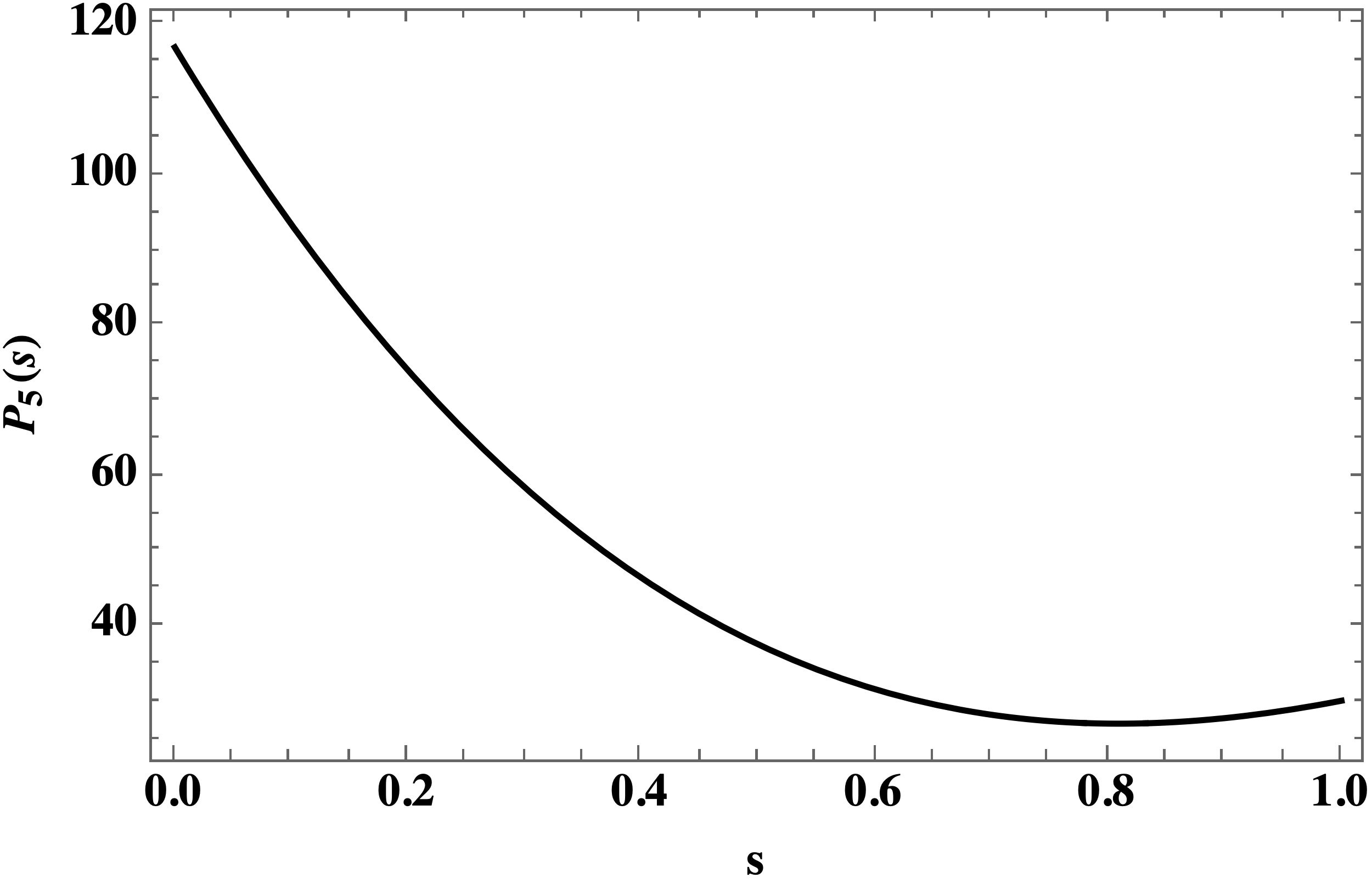}
    \includegraphics[width=0.45\linewidth]{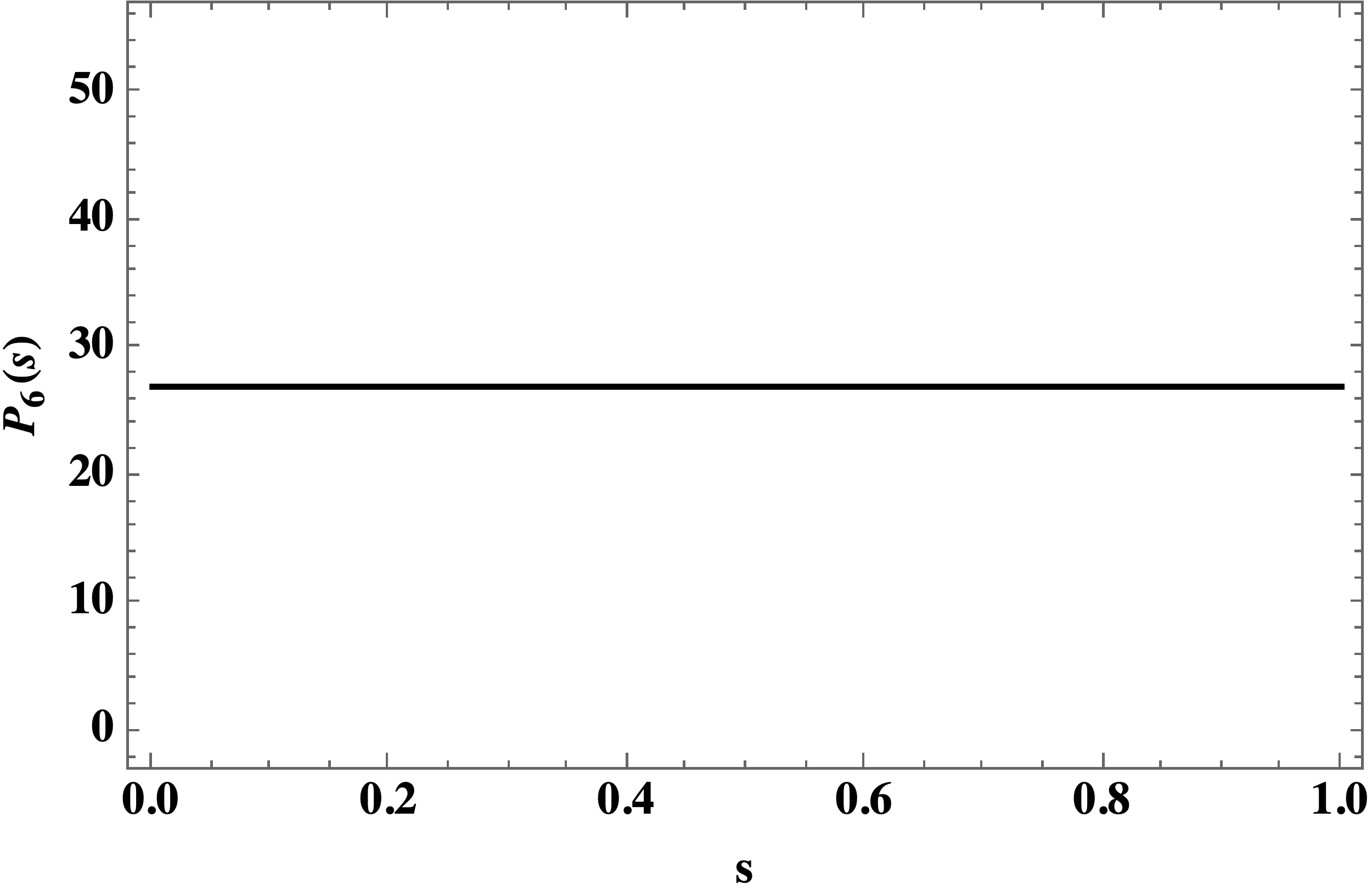}
    \caption{The polynomials \eqref{polys}. }
    \label{polyplot}
\end{figure}

\begin{proof}
    Taking a $\zeta$-derivative in \eqref{defchar} and using \eqref{diffI0} gives
    \begin{equation}\label{defGam}
    \begin{split}
                \frac{\partial\Gamma_{\tau k}(\zeta)}{\partial \zeta} & = \frac{1}{6} \left(6-\zeta ^2+\left(\zeta ^4+\zeta ^2+6\right) k^2 \tau ^2+Z(\zeta ) \left(-\zeta  \left(\zeta ^2-15\right)+\zeta  \left(\zeta ^4+3\right) k^2 \tau ^2\right.\right.\\
                &\quad\left.\left.-2 \ri \left(\zeta ^4-7 \zeta ^2-4\right) k \tau \right)-2 \ri 
   \left(\zeta ^2-2\right) \zeta  k \tau +Z(\zeta )^2 \left(8 \zeta ^2+8 \ri \zeta ^3 k \tau -4\right)\right),
    \end{split}
    \end{equation}
    which is a quadratic polynomial in $Z(\zeta)$. Assume now, to the contrary, that there exists a zero on the imaginary axis $\zeta=\ri y$ such that both \eqref{defchar} and \eqref{defGam} are zero. Eliminating the quadratic term in both expressions leads to 
    \begin{equation}
        \begin{split}
           0&= -\frac{1}{9} (k \tau  y-1) \left(Z(\ri y) \left(3 k^2 \tau ^2+k^2 \tau ^2 y^6+y^4 \left(1-7 k^2 \tau ^2\right)\right.\right.\\
           &\quad\left.+y^2 \left(19 k^2 \tau ^2-4\right)-2 k \tau  y^5+12 k \tau  y^3-18 k \tau  y+5\right)-\\
           &\quad \left. \ri y \left(-7 k^2 \tau ^2+k^2
   \tau ^2 y^4+y^2 \left(1-8 k^2 \tau ^2\right)+2 k \tau  y \left(6 k^2 \tau ^2+7\right)-2 k \tau  y^3-5\right)\right),
        \end{split}
    \end{equation}
    which can easily be solved for $Z(\ri y)$:
    \begin{equation}\label{solZ}
        \begin{split}
            Z(\ri y)=\frac{i y \left(12 k^3 \tau ^3 y+k^2 \tau ^2 \left(y^4-8 y^2-7\right)-2 k \tau  y \left(y^2-7\right)+y^2-5\right)}{k^2 \tau ^2 \left(y^6-7 y^4+19 y^2+3\right)-2 k \tau  \left(y^2-3\right)^2 y+y^4-4 y^2+5}.
        \end{split}
    \end{equation}
    Plugging \eqref{solZ} into either \eqref{defchar} or \eqref{defGam} then gives a rational expression at the assumed degenerate zero, whose numerator is the following polynomial:
    \begin{equation}\label{quotient}
        \begin{split}
            0 & = 6 k^7 \tau ^7 y^3 \left(-y^2-1\right)^2 \left(-y^6+9 y^4-39 y^2-9\right)\\
            &\quad+k^6 \tau ^6 \left(42 y^{12}-311 y^{10}+1031 y^8+2278 y^6+1092 y^4+225 y^2+27\right)\\
            &\quad+2 k^5 \tau ^5 y \left(-63 y^{10}+481 y^8-1654 y^6-2304 y^4-811
   y^2-129\right)\\
   &\quad-k^4 \tau ^4 \left(-210 y^{10}+1615 y^8-5666 y^6-5114 y^4-1436 y^2-117\right)\\
   &\quad-2 k^3 \tau ^3 y \left(105 y^8-795 y^6+2805 y^4+1763 y^2+362\right)\\
   &\quad +k^2 \tau ^2 \left(126 y^8-917 y^6+3221 y^4+1625
   y^2+165\right)\\
   &\quad+2 k \tau  y \left(-21 y^6+143 y^4-499 y^2-245\right)+6 y^6-37 y^4+130 y^2+75. 
        \end{split}
    \end{equation}
    Since any such zero of interest can only occur for $0\leq y<1/(\tau k)$, we set
    \begin{equation}
        s=y/(\tau k),\quad s\in [0,1),
    \end{equation}
    and regroup the numerator of \eqref{quotient} in $s$ to obtain the following polynomial expression of order twelve in s:
    \begin{equation}
        P(s,\tau k)=\frac{1}{(\tau k)^6}\sum_{j=0}^{6} P_j(s) (\tau k)^{2j},
    \end{equation}
    for the polynomials
    \begin{equation}\label{polys}
        \begin{split}
            P_0(s) & = -6 (1-s)^7 s^6,\\
            P_1(s) & = (s-1)^5 s^4 \left(42 s^2-101 s+37\right),\\
            P_2(s) & = -(s-1)^3 s^2 \left(132 s^4-635 s^3+1007 s^2-608 s+130\right),\\
            P_3(s) & = -468 s^7+2278 s^6-4608 s^5+5114 s^4-3526 s^3+1625 s^2-490 s+75,\\
            P_4(s) & = -342 s^5+1092 s^4-1622
   s^3+1436 s^2-724 s+165,\\
   P_5(s) & = 3 \left(-18 s^3+75 s^2-86 s+39\right),\\
   P_6(s) & = 27.
        \end{split}
    \end{equation}
    The polynomials \eqref{polys} are depicted in Figure \ref{polyplot}, which indicates that all expressions are sign-definite except for $P_1$. First, we will now show directly that $P_2,P_3,P_4$ and $P_5$ are sign-definite ($P_0$ and $P_6$ are obvious).\\
    The non-trivial factor of $P_2$ being fourth order with negative discriminant, we can give an explicit expression for its two real zeros which are given approximately as $s\approx 1.59232$ $s \approx 2.21295$ thus proving that $P_2$ has no zeros in $(0,1)$. Similarly, $P_5$ being a cubic polynomial with negative discriminant there is only one real root which can be calculated explicitly and is approximately given as $s\approx 2.68994$. For $P_4$, we note that $P_4(0)=165$, while $P_4(1) = 5$. Since $\partial_sP_4$ is a quartic polynomial with positive discriminant, we can evaluate its zero set explicitly and find that it has no real zeros, while $\partial_sP_4(0)=-724$, which implies that $P_4$ is monotonically decreasing. Thus $P_4$ cannot have any zeros on $(0,1)$. Finally, for $P_3$, we note that $P_3(1)=0$ and we may investigate the sixth-order polynomial $\tilde{P}_3(s)=P_3(s)/(s-1)$. \\
    Now, let us take a closer look at $P_1$. We have the lower bound
    \begin{equation}
    \begin{split}
                P_1(s) & = (-1 + s)^5 s^4 (38 - 101 s + 42 s^2 ) \\
                & \geq (-1 + s)^5 s^4 (38 - 101 s + 42 s^2 - 34/175)\\
                & =: \tilde{P}_1(s).
    \end{split}
    \end{equation}
    Taking a derivative of $\tilde{P}_1$ gives
    \begin{equation}
        \begin{split}
            \frac{d}{ds}\tilde{P}_1(s) & = 2/175 (-1 + s)^4 s^3 (-12882 + 73172 s - 110425 s^2 + 40425 s^3)\\
            & = 2/175 (-1 + s)^4 s^3(s-6/10) (4294 - 17234 s + 8085 s^2),
        \end{split}
    \end{equation}
    with non-trivial zeros given by
    \begin{equation}\label{spm}
        s^{\pm}_0 = \frac{1231 \pm \sqrt{806851}}{1155},
    \end{equation}
    from which only $s^{-}_0$ belongs to $(0,1)$ and, after plugging into the second derivative of $\tilde{P}_1$, reveals that it is the local (and because of the zero structure of $\tilde{P}_1$) global (negative) minimum. It therefore suffices to prove that $P(s^{-1}_0,\tau k)>0$ for all $(k\tau)$. Indeed, we find the sixth-order polynomial in $(\tau k)^2$:
    \begin{equation}\label{pspm}
    \begin{split}
           P(s^{-1}_0,\tau k) & = 27 (\tau k)^{12}+60.0548 (\tau k)^{10}+43.6645 (\tau k)^8+11.7038 (\tau k)^6\\
           &\quad+0.722994 (\tau k)^4-0.01536 (\tau k)^2+0.00032, 
    \end{split}
    \end{equation}
    where the numerical expressions are explicit evaluations of \eqref{spm} to arbitrary arithmetic degree (we show four digits here). Taking a derivative of \eqref{pspm} with respect to $(\tau k)^2$ gives a quintic polynomial. There is only one real zero $(\tau k)_0 $, while the second derivative of $P(s^{-1}_0,\tau k)$  at $(\tau k)_0 $ is non-negative, thus rendering it a local (and, on the real line, global) minimum. Evaluating $P(s^{-1}_0,(\tau k)_0 )>0$ then shows that $P(s_0^{-1},\tau k)>0$ for all $(\tau k)$.\\
 This finishes the proof. 
 \end{proof}

\bibliographystyle{abbrv}
\bibliography{DynamicalSystems}

\end{document}